\documentclass[amsmath,amssymb,pre,12pt]{revtex4}

\usepackage{graphicx}
\usepackage{dcolumn}
\usepackage{bm}
\usepackage{latexsym}
\usepackage{setspace}

\newcommand{\bx}{{\bf x}}

\newcommand{\by}{{\bf y}}
\newcommand{\bp}{{\bf p}}

\newcommand{\bC}{{\bf C}}
\newcommand{\bc}{{\bf c}}

\newcommand{\bB}{{\bf B}}

\newcommand{\CY}{{\cal Y}}

\newcommand{\CM}{{\cal M}}

\newcommand{\bN}{{\bf N}}
\newcommand{\bD}{{\bf D}}

\newcommand{\CK}{{\cal K}}
\newcommand{\bK}{{\bf K}}

\newcommand{\bnabla}{{ \boldsymbol \nabla}}
\newcommand{\bbeta}{{ \boldsymbol \beta}}

\begin{document}

\title{Identification of nonlinear noisy  dynamics of an ecosystem
from observations of one of its trajectory components}

\author
{V.N. Smelyanskiy$^{1\ast}$}

\author
{ D.G. Luchinsky$^{2,1}$}
\author
{M. Millons$^{2}$}

 \affiliation{$^{1}$NASA Ames
Research Center, Mail Stop 269-2, Moffett Field, CA 94035, USA}

\affiliation{$^{2}$Mission Critical Technologies Inc., 2041
Rosecrans Ave. Suite 225 El Segundo, CA  90245}

\email{Vadim.N.Smelyanskiy@nasa.gov}

\date{\today}

\begin{spacing}{1.0}
\begin{abstract}
The problem of determining dynamical models and trajectories that
describe observed time-series data (dynamical inference) allowing
for the understanding, prediction and possibly control of complex
systems in nature is one of very great interest in a wide variety
of fields. Often, however, in multidimensional systems only part
of the system's dynamical variables can be measured. Furthermore,
the measurements are usually corrupted by noise and the dynamics
is complicated by an interplay of nonlinearity and random
perturbations. The problem of dynamical inference in these general
settings is challenging researchers for decades. We solve this
problem by applying a path-integral approach to fluctuational
dynamics \cite{Ludwig:75,Graham:77a,Freidlin:84a,Dykman:90}, and
show that, given the measurements, the system trajectory can be
obtained from the solution of the certain auxiliary Hamiltonian
problem in which measured data act effectively as a control force
driving the estimated trajectory toward the most probable that
provides a minimum to certain mechanical action. The dependance of
the minimum action on the model parameters determines the
statistical distribution in the model space consistent with the
measurements. We illustrate the efficiency of the approach by
solving an intensively studied problem from the population
dynamics of predator-prey system \cite{Hanski:01} where the prey
populations may be observed while the  predator populations or
even their number is difficult or impossible to estimate. We
emphasize that the predator-prey dynamics is fully nonlinear,
perturbed stochastically by environmental factors and is not known
beforehand (see e.g.~\cite{Clark:03}). No overall solution was
previously available for this problem even in the deterministic
case~\cite{Kurths:04,Wood:01}. We apply our approach to recover
both the unknown dynamics of predators and model parameters
(including parameters that are traditionally very difficult to
estimate) \emph{directly} from measurements of the prey dynamics.
We  provide a comparison of our method with the Markov Chain Monte
Carlo technique. As a further test of the method we demonstrate
the reconstruction of the dynamics of chaotic Lorenz attractor
driven by noise from measurements of only one if its trajectory
component.
\end{abstract}
\end{spacing}
\maketitle

\section{Introduction}

For quantitative understanding, predicting, and controlling
time-varying phenomena it is necessary to relate observations to a
mathematical model of a system dynamics. In a great number of
important problems such  model is multidimensional, nonlinear,
stochastic and not known from ``first principles''. Furthermore,
often only part of the system's variables can be measured and
these measurements are corrupted by noise. The rest of the system
variables are invisible, or \textit{hidden}. In these settings,
perhaps the most fundamentally difficult unsolved problem of
dynamical inference is how and to which extent one can learn both
the model parameters and system trajectory from a given set of
incomplete trajectory measurements.  A solution of this problem is
of importance across many disciplines. Examples range from
molecular motors~\cite{Visscher:99} to coupled matter-radiation
system~\cite{Christensen:02} (see e.g.
~\cite{Abarbanel:01,Kurths:04,Wood:01,Hanski:97} for further
examples).

Here we present a solution to this problem using a path-integral
approach to fluctuational dynamics\cite{Ludwig:75}. We show that,
given the measurements, the most probable system trajectory can be
obtained from finding the minimum of the mechanical action of a
certain auxiliary Hamiltonian system under properly defined
boundary conditions. The dependence of the minimum action on the
system model parameters determines the statistical likelihood of
different parametric models.

\begin{figure}[t!h]
 \begin{center}
 \includegraphics[width=10cm,height=5cm]{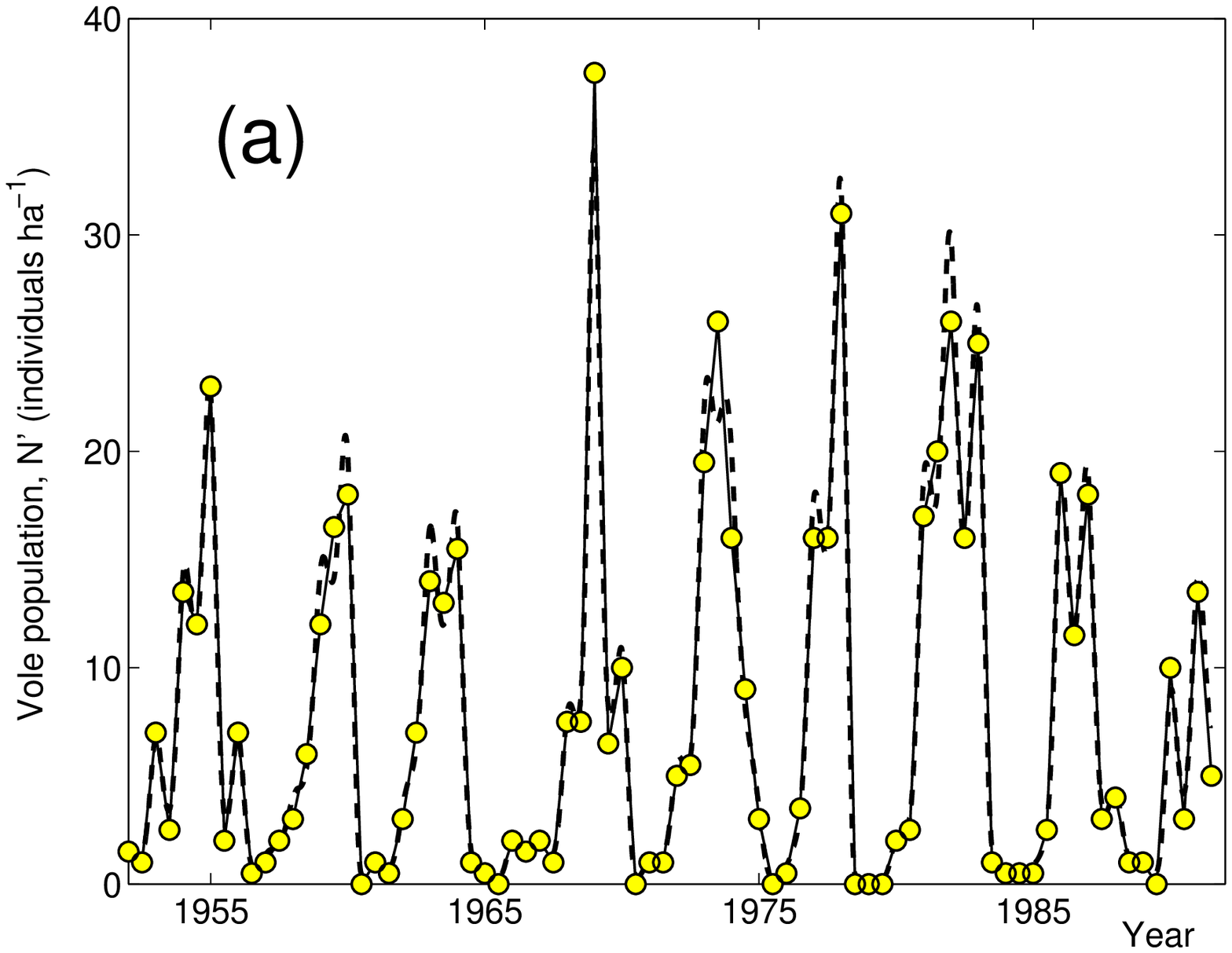}
 \includegraphics[width=10cm,height=5cm]{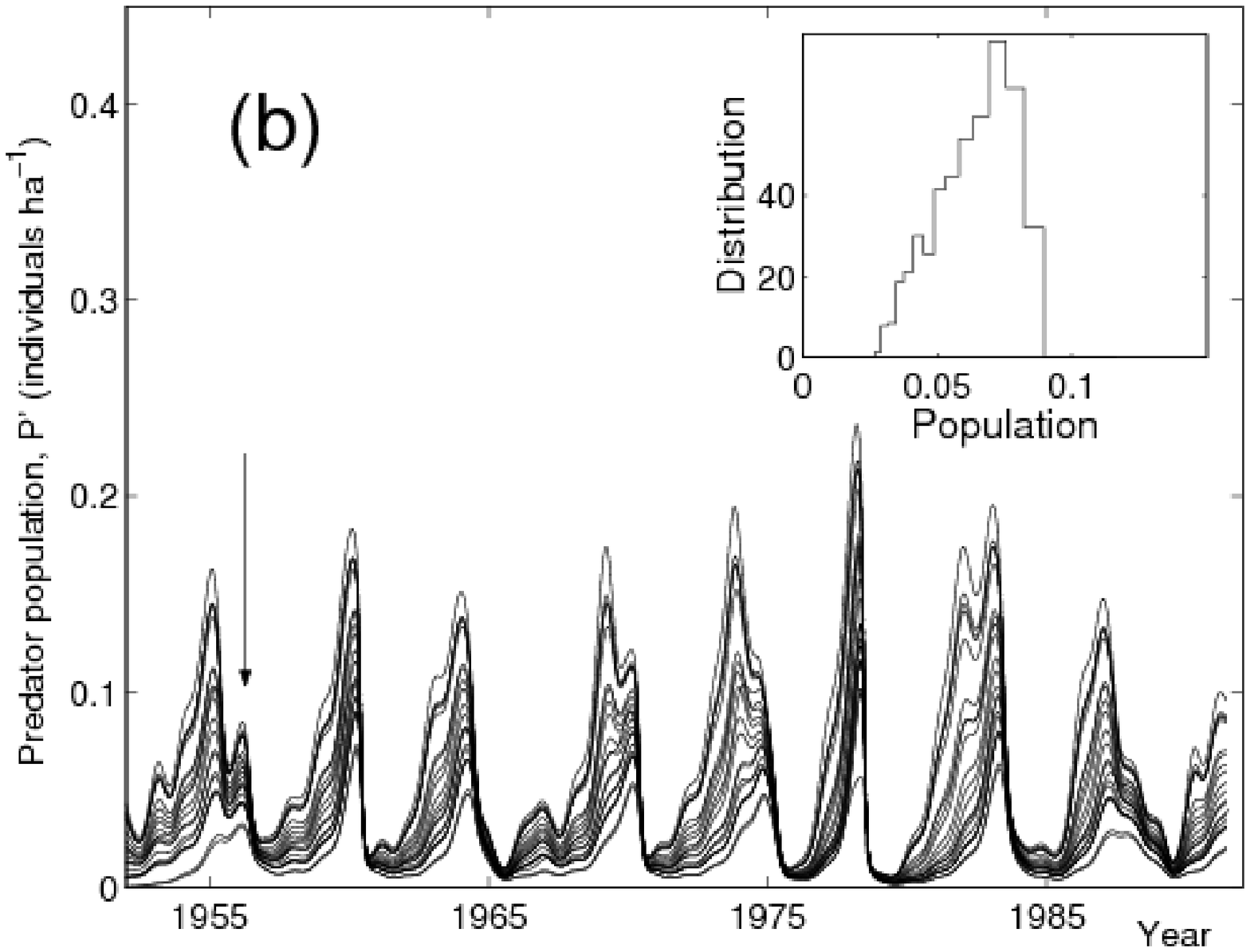}
 \end{center}
 \caption{\label{fig:vole_original} (a) Stochastic trajectory of the
population dynamics of small rodents observed in
Kilpisj$\ddot{a}$rvi, Finnish Lapland, 1952-1992~\cite{Hanski:97}
is shown by yellow dots. Black solid line is shown to guide an
eye. Dashed lines shows the solution of the optimization problem.
(b) Black solid line shows recovered hidden dynamics of the
population of the specialists predators obtained by varying
parameters $r$ and $rK'/K$ of the model (\ref{eq:the_model}).
Parameters used to obtain these results are: r  = 5.2 $\pm$ 2.5,
rK'/K = -5.2 $\pm$ 2.5; s = 1.2; a = 15; g  = 0.1; e1 = .8; e2 =
.5; K = 90; Q = 30; $\sigma_{n}$ = 0.02; $\sigma_{p}$ = 0.02. The
insert shows the cross-section of the weighted distribution of the
dynamical trajectories for the year 1956 indicated by the arrow in
the main figure.}
 \end{figure}

The method is used to solve an intensively studied problem from
the population dynamics of the predator-prey
system\cite{Hanski:97,Hanski:01,Turchin:00} where the cyclic
dynamics of populations of small rodents is observed in
Kilpisj$\ddot{a}$rvi, Finnish Lapland, 1952-1992 \cite{NERC} (see
Fig.~\ref{fig:vole_original}(a)) while the number of predators is
difficult or impossible to estimate. The predator-prey dynamics is
fully nonlinear subject to seasonal and random perturbations. This
is a classical longstanding problem in
ecology~\cite{Volterra:1926} and epidemiology (see
e.g.~\cite{Schwartz:84}). In particular, the cited database
accumulates nearly 5000 individual datasets with similar structure
collected over more then 150 years of research. It is shown that
the proposed approach allows to recover both the unknown dynamics
of predators and model parameters directly from measurements of
the prey dynamics.

\section{Path-integral approach to the problem of dynamical
inference.}

To formalize the discussion above note that in a typical
experimental situation we observe $M$-dimensional time series of
signals $\CY\equiv\{{\bf y}(t_n) = {\bf y}(t_0+hn),\,~
n=\overline{0,\CK}\}$, with the sampling step $h$. The unknown is
the actual $L$-dimensional  dynamical trajectory of the system
${\bf x}(t)$. We are interested in the case where $M\leq L$ so
some of the trajectory components are hidden.
Quantitative understanding of the time-varying phenomena
underlying $\CY$ requires, in general, an expert input into the
observed data in the form of a mathematical modelling framework
for the system dynamics and for the measurement scheme. A commonly
used dynamical and measurement equations for nonlinear models in
the presence of random perturbations  that apply to but by far not
limited by the predator-prey ecological system described above are
 \begin{eqnarray}
 \label{eq:SDE}
 &&\dot x_i(t) = K_i(\bx(t),{\bf c})+ \xi_i(t) , \quad
  \langle \xi_i(t)\, \xi_j(t')\rangle = D_{ij}\delta(t-t'),  \\
 && y_k(t) =\sum_{i=1}^{L}B_{ki}\, x_i(t)+ \beta_k(t),\quad
 \langle\beta_k(t) \,\beta_{l}(t^{\prime})\rangle = N_{k l}\,\delta(t-t^{\prime}).
\end{eqnarray}\noindent
Here we introduced  continues-time interpolations $y_k(t)$ for
components of the observed time-series $\CY$. This approximation
can often be justified for a sufficiently small sampling interval
$h$  and a large number of data points, $\CK\gg 1$, in the
time-series $\CY$. In (\ref{eq:SDE}) $x_i(t)$ ($i=1:L$) are
dynamical variables composing a vector
 $\bx(t)$ that describes an instantaneous state of the system.
  The system dynamics in (\ref{eq:SDE}) is governed by
$L$-dimensional vector field with components $K_i$ depending on
the set of parameters $\{c_\alpha\}\equiv \bc$ and white Gaussian
process with zero-mean components $\xi_i(t)$ characterized by a
$L\times L$ correlation matrix $\hat {\bf D}$. The deterministic
part of the measurement equation in (\ref{eq:SDE}) is described by
$M\times L$ measurement matrix $\hat {\bf B}$ and the measurement
error is described  by the white Gaussian process with zero-mean
components $\{\beta_m(t)\}$ and $M\times M$ measurement noise
matrix $N_{n\,m}$ ($M\le L$). Overall,  the dynamical and
measurement model (\ref{eq:SDE}) is characterized by the full set
of the \emph{unknown} parameters $\{\CM_\alpha\}\equiv\{\bc,
\hat{\bf B}, {\bf\hat D}, {\bf\hat N}\}$.

 Due to the
presence of  dynamical and measurement noise the problem of
dynamical inference  must be cast in probabilistic terms. This can
be done within a general framework of the Baeysian statistical
approach ~\cite{Congdon:01,Meyer:01}. A key statistical quantity
is a so-called {\it likelihood} probability density functional
(LPDF) ${\cal P}_{\CY}[{\bf x}(t); \CM]$. It represents a joint
probability density  that the system trajectory is $\bx(t)$ and
the system parameter values are $\{\CM_\alpha\}$ conditioned on
the observed time-series ${\cal Y}$.  We emphasize that in a real
physical process the system has a distinct trajectory and
parameter values. In this regard the LPDF represents a degree of
uncertainty in our knowledge about these quantities obtained from
the measurements and assuming some basic properties of the system
fluctuational dynamics \cite{prior}.

The explicit form of the LPDF  can be obtained from (\ref{eq:SDE})
using the path-integral approach to fluctuational dynamics
~\cite{Ludwig:75,Graham:77a,Dykman:90}. We write    ${\cal
P}_{\CY}[{\bf x}(t);\CM]=A_{\CY} \exp\left(-S_{ {\cal Y}}[{\bf
x}(t);\,\CM]\right)$, where $A_\CY$ is a normalization constant
and a negative log-likelihood functional $S_\CY$ is obtained in
the Appendix A
\begin{eqnarray}
 \label{eq:action}
 &&S_{{\cal Y}}[{\bf x}(t); \CM]=
 \frac{1}{2}\int_{0}^{T}dt
 \left[\left(\by(t)-\hat \bB \,\bx(t) \right)^T{\bf\hat
 N}^{-1}\left(\by(t)-\hat \bB\,\bx(t) \right)+\bnabla\cdot
 \bK(\bx(t),\bc)\right.\nonumber\\&&
\left.+\left(\dot\bx(t)-\bK(\bx(t),\bc)\right)^T\hat
 {\bf D}^{-1}(\dot\bx(t)-\bK(\bx(t),\bc))\right]+\frac{\CK}{2}\ln\left(\det(\hat{\bf D})\det(\hat{\bf N})\right).
\end{eqnarray}
\noindent Here $T=\CK h$ is a time length of the data record
$\CY$. In what following we shall focus on the case $M < L$ that
implies the existence of hidden variables. We note that despite
hidden dynamical variables are not measured directly the
functional $S_\CY$ (\ref{eq:action}) still depends on them
explicitly because of the dynamic coupling between the variables
imposed by the force field $\bK$.

In many practically important cases available recording of a
system trajectory,  while containing  only a part of dynamical
variables, has sufficiently small time-step, long time duration
and  limited noise characteristics. Such measurements can provide
a strong information that is sufficient to pin down both key model
parameters of the system and its trajectory, or at least to
extract strong correlations between  them. In these cases  the
joint LPDF $P_{\CY}(\bx(t),\CM)$ will be well localized in the
vicinity of one or more of its maxima where $\delta
S/\delta\bx(t)=0$ and $\{\partial S/\partial \CM_\alpha=0\}$. In
the
 case where a single maximum dominates LPDF its position
corresponds to the trajectory $\bx^{\rm opt}(t)$ and parameter
values $\{\CM_{\alpha}^{\rm opt}\}$ that the system most probably
has, given the measurements $\CY$.

We now put forth  a new paradigm in which a  solution of the
dynamical inference problem with hidden variables is obtained via
the calculus of variations for the functional $S_\CY(\bx(t),\CM)$.
 The power of this approach is in its simplicity,
efficiency and an insight that it provides  to the solution of
dynamical inference problem by drawing a close connection to the
methods and concepts of classical mechanics, in particular, a
least action principle.

We search for the  minimum of $S_\CY(\bx(t),\CM)$ by alternatively
computing the expected values of $\bx(t)$ and model parameters in
$\CM$
 from the solution of the two variational problems
$\frac{\delta S}{\delta \bx(t)}=0$ and $\frac{\delta S}{\delta
\CM}=0$.
 The first condition  corresponds to a solution of the boundary value problem
 for an auxiliary mechanical system with
 the coordinate ${\bf x}$,
 momentum ${\bf p}$ and a Hamiltonian function $H(\bx,\bp)$
\begin{eqnarray}
 \label{eq:the_Hamiltonian}
&& H(\bx,\bp)=-\frac{1}{2}\left(\by-\hat\bB\, {\bf
x}\right)^T{\bf\hat
 N}^{-1}\left(\by-\hat\bB \,{\bf x}\right)-\frac{1}{2} \frac{\partial \bK}{\partial \bx}+\bK
 \bp
+\frac{1}{2}\bp^T\,\hat
{\bf D}\,\bp, \\
&&\bp=\hat {\bf D}^{-1}(\dot\bx(t)-\bK).
\end{eqnarray}
\noindent  We  look for the solution of the Hamiltonian equations
\begin{eqnarray}
 \label{eq:HE}
&& \dot \bx =\bK + \hat{\bf D}\bp, \qquad \dot
\bp=\frac{1}{2}\frac{\partial^2 \bK}{\partial
\bx^2}-\frac{\partial \bK}{\partial \bx}\bp -\left(\by-\hat \bB\,
{\bf x}\right)^T{\bf\hat
 N}^{-1}\hat {\bf B}
\end{eqnarray}
\noindent that satisfy  the  boundary conditions
$\bp(0)=\bp(t)=0$. If several solutions exist we choose the one
providing a minimum of a functional $S_{{\cal Y}}[{\bf x}(t);
\CM]$ playing a role of a mechanical action.
 We then fix the inferred trajectory $\bx(t)$ and update the
value of the parameters in the set $\CM$,
 using analytical solution
of the second variational problem, $\frac{\delta S}{\delta
\CM}=0$, developed in our earlier research~\cite{Smelyanskiy:05b}
(see Appendix A for  details). This procedure is repeated
iteratively until the desired convergence is achieved. The outcome
of this algorithm is the most probable system trajectory  ${\bf
x}^{\rm opt}(t)$ and  model parameters ${\cal M}^{\rm opt}$. The
measure of their fitness to the observed data ${\cal Y}$ is
$\propto \exp\left(S^{\rm opt}\right)$ where
 the globally minimum action $S^{\rm
opt}=S_{ {\cal Y}}[{\bf x}^{\rm opt}(t);\,{\cal M}^{\rm opt}]$.

The  Bayesian  approach for dynamical inference  was initially
proposed by Meyer and Christensenin \cite{Meyer:01} for the case
where all variables were directly observed. The previous work on
this subject (see
e.g.~\cite{Meyer:01,Rossi:02a,Clark:03,Friedrich:03a})  was
exclusively focusing on brute force  numerical methods, such as
Markov Chain Monte Carlo (MCMC).  However our detailed study of
MCMC approach for the problem of dynamical inference with hidden
variables has shown that the functional $S_{\CY}(\bx(t),\CM)$
 has multiple deep spurious minima in the space of piece-wise continuous trajectories
$\{\bx(t_m);\,m=1:M\}$. These minima occur  due to the
contributions to the cost functions from the  terms  of the order
of $\frac{(x_{i}(t_{k+1})-x_{i}(t_{k}))^2}{h^2}$. If one starts
from a  poor guess about both the system trajectory and model
parameters MCMC search stacks in those minima and takes a
prohibitively large time to converge (see Sec.~\ref{sec:MCMC} for
details). In contrary, our approach avoids those  spurious minima
because the solution of the Hamiltonian boundary value problem
(\ref{eq:HE}) is achieved via large smooth variations in the space
of continuous trajectories. This key finding reflects a basic
property of a hidden-variable and parameter inference in a noisy
dynamical system with continuous vector field $\bK(\bx,\bC)$:
expected value of the inferred trajectory $\bx^{\rm opt}(t)$ is a
smoothly varying function of time whereas the measured signal
$\by(t)$ is not.

We note that the likelihood distribution around the maximum is
determined by the second variation $\delta^2 S_\CY$ of the action
with respect to the both ${\bf x}(t)$ and $\CM$ computed at its
minimum. In many cases, in particular in the case of a multi-modal
LPDF, it is of interested to explicitly study the full shape of
the LPDF in a reduced subspace of the model parameters while
marginalizing the LPDF with respect to the other parameters and
the system trajectory. Within our approach this can be handily
done  by computing the minimum action $S_{\rm opt}(c_1,c_2)$ using
the above algorithm for different values of the parameters
$(c_1,c_2)$.

However in many complex cases where observational and model errors
are significant and hidden variables are present LPDF can have a
very large number of local minima. In this case the more
informative quantity is a distribution of local minima. To obtain
it we pick at random some values of ($c_1,c_2)$ and converge to
the nearest point ($c_{1}^{\prime},c_{2}^{\prime}$) of a local
minimum of the action $S_{\rm opt}(c_1,c_2)$ where the conditions
$\frac{\delta S}{\delta \bx(t)}=0$ and $\frac{\delta S}{\delta
\CM}=0$ are satisfied. We then repeat this procedure many times
for different starting values of ($c_1.c_2)$. Then the histogram
of the local minima ($c_{1}^{\prime},c_{2}^{\prime}$) weighed with
the factors $\exp(-S_{\rm opt}(c_{1}^{\prime},c_{2}^{\prime})$ and
appropriately normalized gives the distribution of local minima
$P=P_{\cal Y}(c_1,c_2)$. We will demonstrate this approach in
Sec.~\ref{sec:predator.prey} for the inference  of the population
dynamics.

Finally, we emphasize that the  prerequisite of the approach
considered in this section is that LPDF computed at any set ${\cal
M}$ of relevant parameter values has a sharp peak in the space of
the system paths $\bx(t)$  at some ${\bf x}^{\rm opt}(t)$ that
depends on ${\cal M}$ (cf.  Fig.~\ref{fig:3dtrajectory}).
\begin{figure}[t]
\begin{center}
\includegraphics[width=10cm,height=6cm]{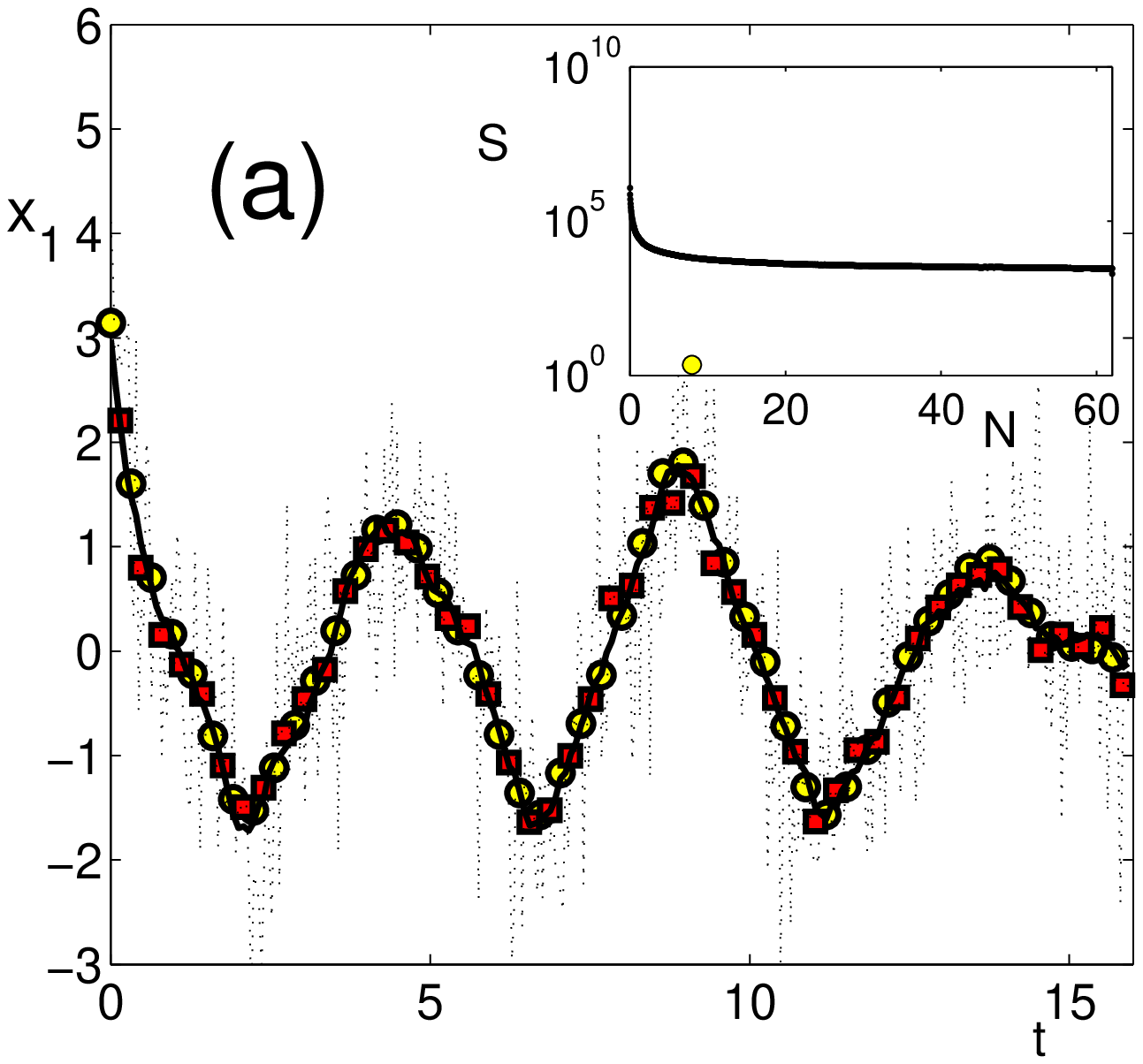}

\includegraphics[width=10cm,height=6cm]{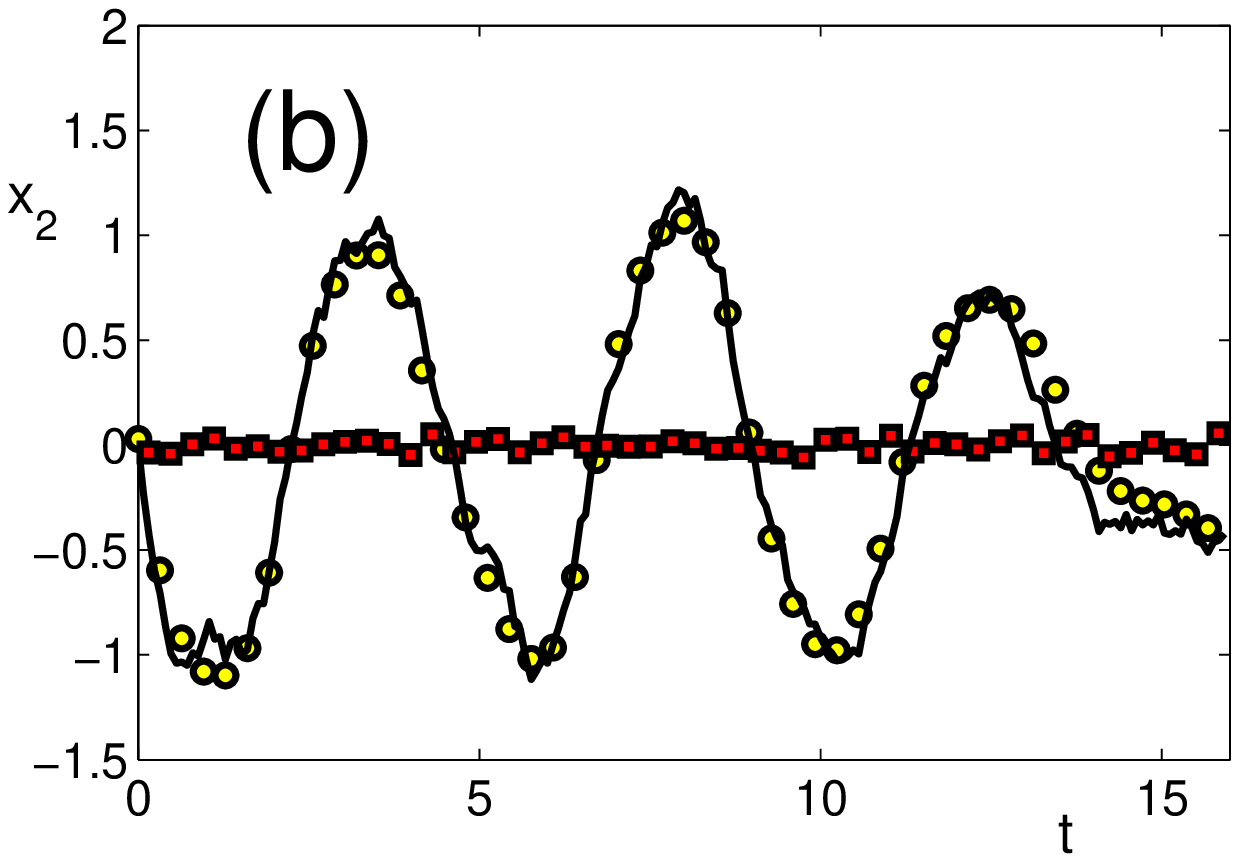}
\end{center} \caption{\label{fig:bvp_mcmc} Result of the direct comparison
of the path-integral and MCMC techniques for (a) observed variable
$x_1(t)$ and (b) hidden variable $x_2(t)$. The actual dynamical
trajectories $x_1(t)$ and $x_2(t)$ are shown by solid lines. The
measured trajectory $y_1(t)$ is shown by dashed black line in
figure (a) and is taken as initial guess for the solution
$x_1(t)$. For an unobservable trajectory $x_2(t)$ initial guess is
taken to be $y_2(t)=0$. The solution of the boundary value problem
is shown by yellow circles. The MCMC solution is shown by red
squares. The inset in the figure (a) shows the variation of the
cost function as a function of time for MCMC algorithm (black
dots) and for boundary value method (yellow circle).}
\end{figure}

\section{\label{sec:predator.prey} Inference of predator-prey model.}

We now apply the method described above to reconstruction of the
unknown predator dynamics and model parameters from the observed
oscillations of small rodents in Finish Lapland. The observed
time-series data is shown by yellow circles in the Fig. 1(a). To
formulate the problem we first briefly summarize an expert input
into observed data, see e.g.
~\cite{Hanski:97,Turchin:00,Hanski:01,Hanski:03} for more details.
It was argued~\cite{Hanski:97,Hanski:01} that the most likely
predators to potentially maintain oscillatory dynamics in rodent
populations are small mustelids, weasels, and stoats which are
notoriously difficult to observe and study in the field. It was
further argued~\cite{Hanski:97,Hanski:01,Hanski:03} that in
addition to the dominating effect of these so-called specialist
predators the population of rodents is strongly affected by
generalists predators (such as foxes, owls and skua) and by
seasonal (periodic) and stochastic variations of the environment.
Based on these arguments  the following equations were introduced
to model observed ecological time series
\begin{eqnarray}
    \label{eq:the_model}
  &&\dot{N} = r N \left(1-e_1\sin(2\pi t)+\sigma_n\xi_n(t)\right)
   - (r/K)N^2  - \frac{GN^2}{N^2+H^2} - \frac{CNP}{N+D}, \nonumber \\
  &&\dot{P} = s P \left(1-e_2\sin(2\pi t)+\sigma_p\xi_p(t)\right) -
  sQ\frac{P^2}{N}.
\end{eqnarray}
Here the  state of the system is characterized by the dynamical
variables $N$ and $P$, corresponding to the density of rodents and
predators, respectively. Taking into account a log-normal
distribution of the measurement errors the measured rodent density
 $N'$  is related to the actual (unknown)
value $N$ via $N' = N\exp(\sigma_{obs}\eta(t))$ where $\eta(t)$ is
a white Gaussian noise of unit intensity. The predator density is
not measured so the variable $P$ is hidden. In
(\ref{eq:the_model}) $\{\xi_n(t), \xi_p(t)\}$ is a zero-mean white
Gaussian vector of dynamical noise.  The precise functional form
is known neither for predation nor for numerical response of the
predators and some modifications of the equations
(\ref{eq:the_model}) where considered in the
literature~\cite{Turchin:00}.

The problem of dynamical inference is the following: Use 80
experimental points of corrupted by noise measurements to recover
both hidden dynamics of predators $P=P(t)$ and the model of the
nonlinear stochastic dynamics of small rodent in Fennoscandia
represented by the full set of parameters from
Eq.~(\ref{eq:the_model}) and $\sigma_{osb}$.  Since there were no
general methods to recover neither hidden dynamics nor nonlinear
models of stochastic systems it was always assumed (see
e.g.~\cite{Turchin:00,Hanski:01}) that the goal to obtain solution
of this problem is unrealistic and no attempt was made to solve it
in the earlier research. Instead a number of models were
developed~\cite{Hanski:97,Turchin:00,Hanski:01,Hanski:03} from the
first ecological principles and from the extensive field studies
of the small rodents ecology. The outcome of the simulation of
these models was compared to the experimental points to decide
whether or not the model is capable of producing reasonable
predictions. This approach although very valuable and often the
only one available in practice has very limited statistical
significance and can hardly be generalized.
\begin{figure}[ht!]
\begin{center}
\includegraphics[width=6cm,height=6cm]{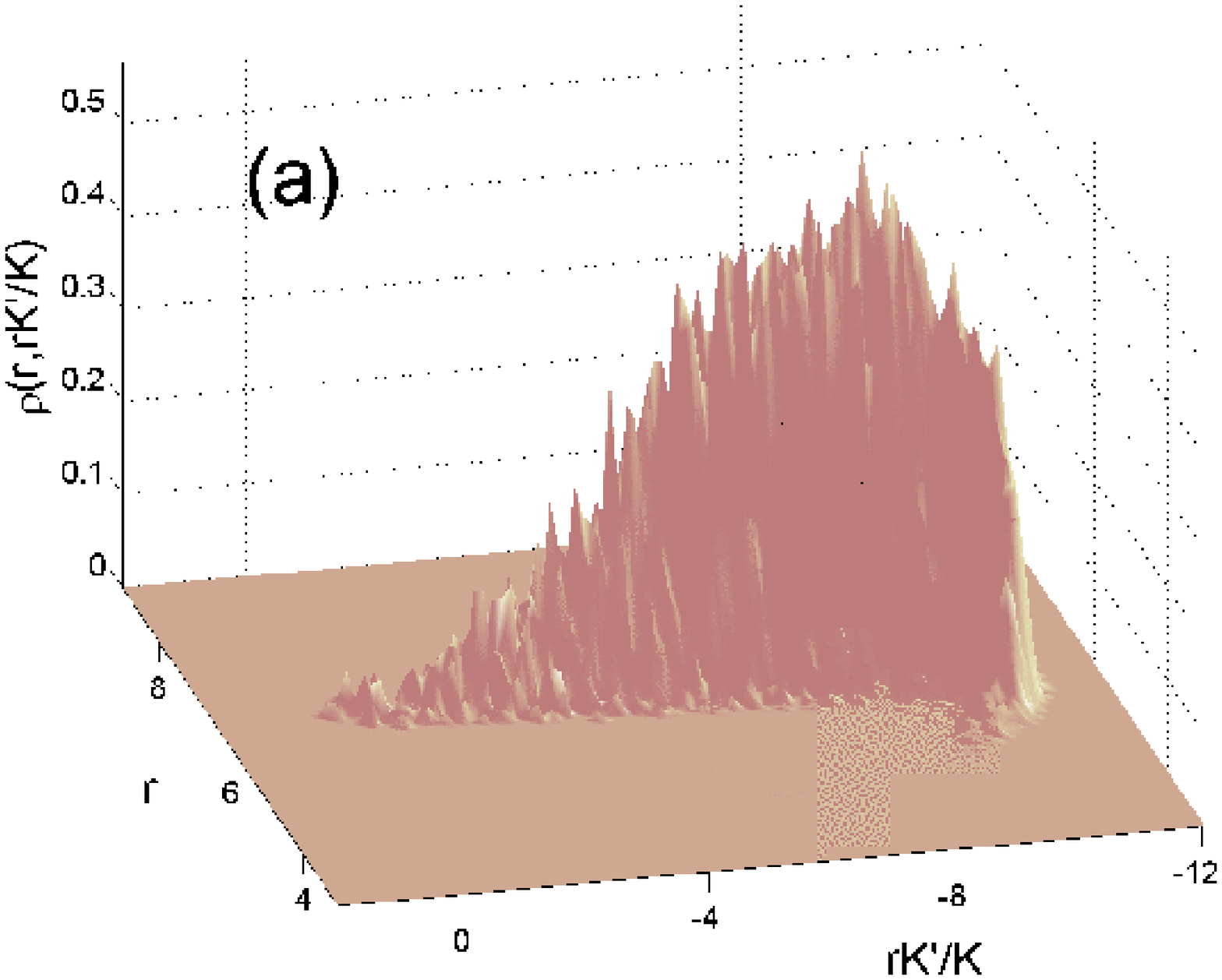}
\includegraphics[width=6cm,height=5.5cm]{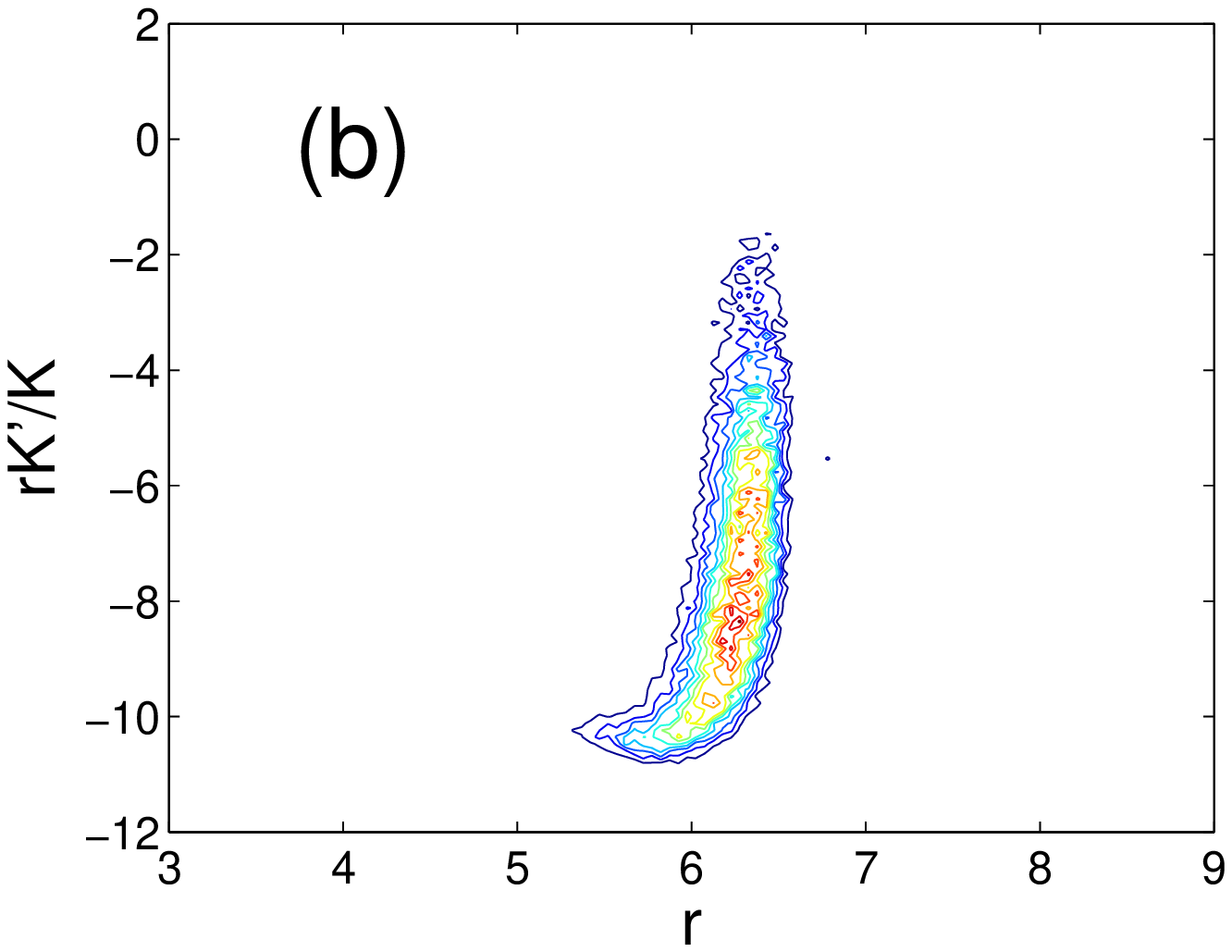}

\includegraphics[width=6cm,height=6cm]{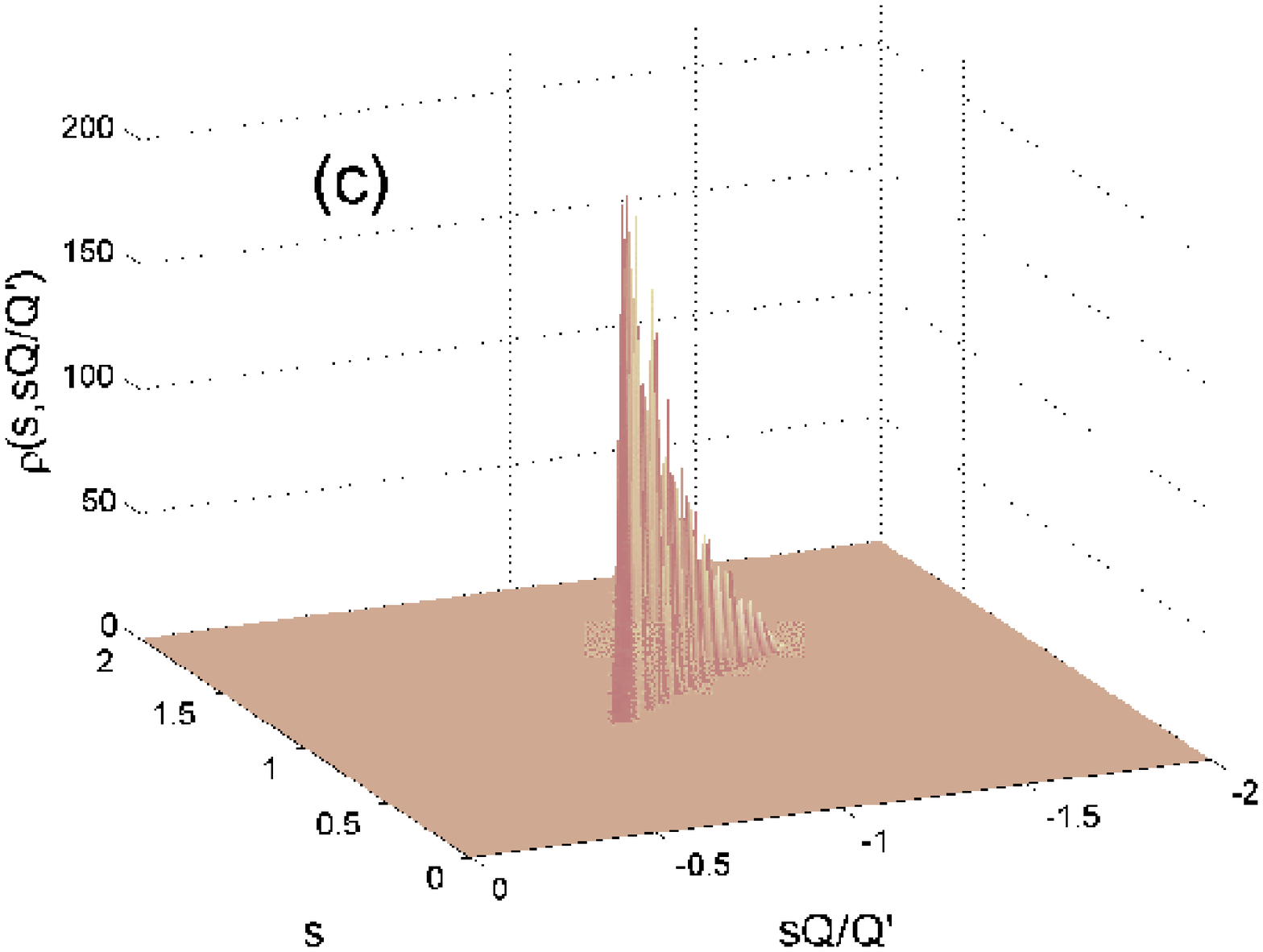}
\includegraphics[width=6cm,height=5.5cm]{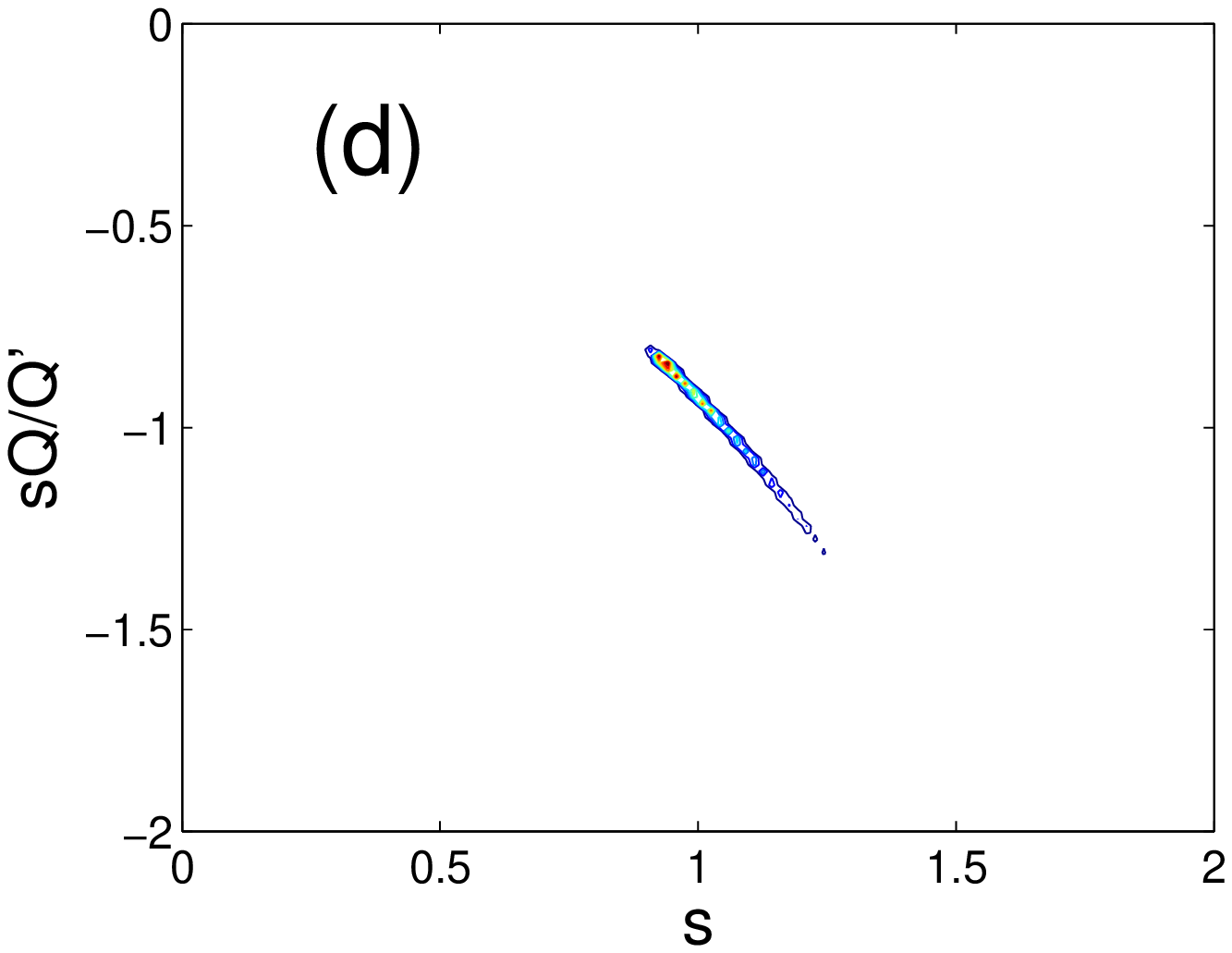}
\end{center} \caption{\label{fig:vole_rs_dist} (a) Weighted
distribution of the inferred values of the model parameters $r$
and $rK'/K$. (b) The same distribution top view. (c) Weighted
distribution of the inferred values of the model parameters $s$
and $sQ/Q'$. (d) Top view of the same distribution.}
\end{figure}

The method introduced above provides a general and effective
alternative approach to a solution of this ecological problem.
First, we map the predator-pray model (\ref{eq:the_model})
directly onto the dynamical model with additive white noise
considered in Eqs.~(\ref{eq:SDE}) by making the change of
variables: $x_1(t)=\log (N(t)/K')$ and $x_2(t)=\log(Q' P/K')$
(here some known nominal values are used for the scaling
coefficients $K'$ and $Q'$). Then the  full set of unknown
parameters  $\bc =
\{r,s,e_1,e_2,K,G,C,Q,H,D,\sigma_n,\sigma_p,\sigma_{obs}\}$ and
the  trajectory of the predator density $P=P(t)$ is inferred from
the observed data using the dynamical inference scheme described
above (for the scaling of dynamical equations and precise
ecological meaning of these parameters
see~\cite{Hanski:97,Hanski:01,Hanski:03}). To present the solution
of this inference problem we investigate the marginalized LPDF as
a function of  key model parameters
 that are notoriously difficult to
estimate~\cite{Hanski:97} using other techniques: carrying
capacity $K$ and equilibrium ration between two populations $Q$
(see online supplement material for further details).

The results are shown in the Fig~\ref{fig:vole_original} and
Fig.~\ref{fig:vole_rs_dist}. It can be seen from the
Fig~\ref{fig:vole_original} (a) that the model
(\ref{eq:the_model}) can fit experimental data very well in a wide
range of values of e.g. parameters $r$ and $rK'/K$. This gives
rise to a broad distribution of the possible dynamical
trajectories of hidden predators shown in the
Fig~\ref{fig:vole_original} (b). However, the likelihood functions
of various trajectories are exponentially different. This fact is
taken into account by weighting the corresponding distributions of
the model parameters with the factor
$\exp(-S_\CY\left[\bx(t),\bc\right])$. The weighted distributions
of trajectories and model parameters is the main outcome of the
statistical analysis of the ecological experimental data.

The weighted joint distributions of the inferred pairs of
parameters ($r$, $r/K$) and ($s$, $sQ/Q'$) are shown in the
Fig.~\ref{fig:vole_rs_dist}. Analysis of these distributions gives
the following estimates of the model parameters $r=5.69\pm 0.49$,
$rK'/K=-6.0494\pm 1.25$, $K= 76\pm  17$ $s=1.08\pm 0.31$,
$sQ/Q'=-1.17\pm 0.50$, $Q= 43\pm 22$, $g = 0.12 \pm 0.3$, $a =
13.2 \pm 2.5$, $e_1=1.4\pm .4$, and $e_2=1\pm .5$ which are close
to the values considered in the earlier ecological
research~\cite{Hanski:97,Turchin:00,Hanski:01,Hanski:03}. At the
same time statistical analysis reveals that distributions for the
parameters $\{H, D\}$ (in the range of values ???) are very flat
and further information is needed for a more accurate estimate of
their values.

\section{Lemming oscillations in the high-arctic tundra in
Greenland}
 \label{s:lemming}

The method can be further verified by analyzing experimental data
obtained for the small rodents-predators community in high-arctic
Greenland~\cite{Hanski:03}. This data is very similar to the data
collected in Fennoscandia with a important exception, namely,
dynamics of both populations prey and predator was recorded very
carefully in Greenland. Therefore, it has become possible to check
if the predator dynamics reconstructed from the prey population
alone coincides with the actual observations of the predator time
series.
\begin{figure}[ht!]
\begin{center}
\includegraphics[width=10cm,height=5cm]{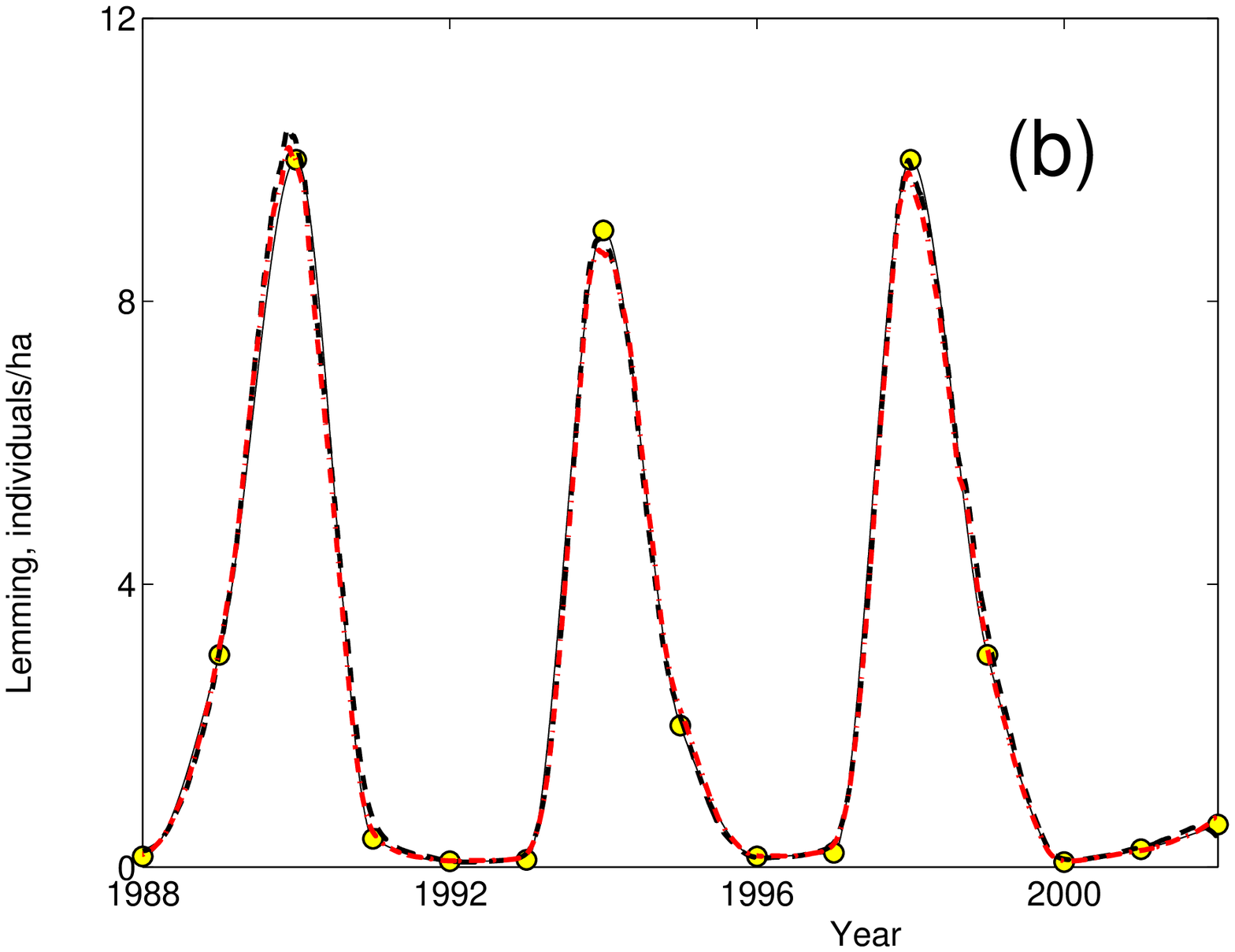}

\includegraphics[width=10cm,height=5cm]{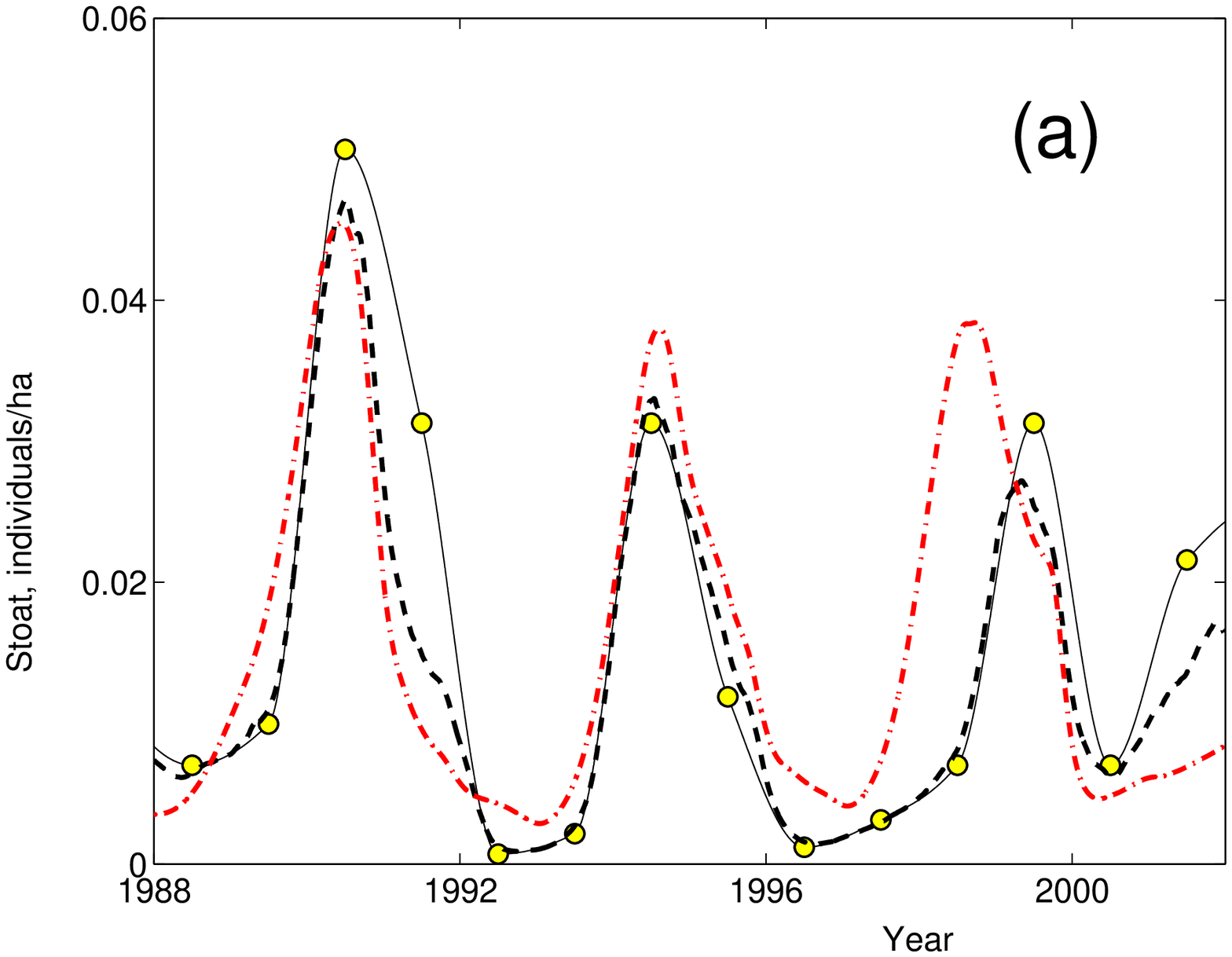}
\end{center} \caption{\label{fig:lemming_data} Lemming (a) and (b)
stoat population (individuals/ha) observed in the high-arctic
tundra in 1988--2002~\cite{Hanski:03} are shown by yellow circles.
The thing solid line is shown to guide the eye. Dashed black lines
show population dynamics inferred using model
(\ref{eq:lemming_model}) under assumption that the dynamics of
both populations was measured in the experiment with measurement
error 0.2. Dashdot lines show population dynamics inferred using
model (\ref{eq:lemming_model}) under assumption that only the
dynamics of prey populations was measured in the experiment with
measurement error 0.2 and the predator dynamics is hidden.}
\end{figure}
The experimental data (see Fig. 3) for the population oscillation
in the lemming-stoat predator prey community were collected in the
high-Arctic tundra in Greenland in 1988--2002~\cite{Hanski:03}.
The time-variation of the predator and prey oscillations are also
influenced by a number of generalists predators such as arctic
fox, snowy owl, and long-tailed skua. We note that a very detailed
model with experimentally measured numerical responses of various
predators is available~\cite{Hanski:03} to simulate this data.
However, to analyze hidden predator population along the lines
outlined in the previous section we notice that the organization
of the lemming-stoat community in the high-arctic Greenland is
very similar to the vole-weasel community in Finnish Lapland. So
we attempt to fit the lemming-stoat population oscillations using
the model (\ref{eq:the_model}) developed for the latter
community~\cite{Hanski:97,Turchin:00,Hanski:01}.

In this model the populations  are scaled as $x=log(N/K')$ and
$y=log(\frac{Q'P}{K'})$ with some assumed values of the carrying
capacity $K'$ and proportionality constant $Q'$ ($K'$ and $Q'$ are
known, while actual values $K$ and $Q$ are not known and have to
be inferred). The time-variations of $x(t)$ and $y(t)$ are
described by the following set of equations~\cite{Hanski:97}
\begin{eqnarray}
    \label{eq:lemming_model}
  &&\dot {x} = r \left( 1 -e_1\sin(2\pi t)\right) - r\frac{K'}{K}e^x  - g\frac{e^x}{e^{2x}+h^2} - a\frac{e^x}{e^2+d}, +D_n\xi_n(t)\nonumber \\
  &&\dot {y} = s \left( 1 -e_2\sin(2\pi t)\right) - s\frac{Q}{Q'}e^{y-x}+D_p\xi_p(t).
\end{eqnarray}
The parameters of these equations have the
following~\cite{Hanski:97} meaning. The {\it vole population} is
characterized by: (i) the intrinsic rate of the vole population
growth $r$ with possible values in the range: 4 - 7 yr$^{-1}$;
(ii) the dimensionless amplitude of seasonal forcing $e$ with
range: 0.5 - 1; (iii) prey carrying capacity $K$ with range: 100 -
300 voles ha$^{-1}$. The {\it specialist predator population} is
described by: (i) intrinsic rate of weasel population growth $s$
with range: 1- 1.5 yr$^{-1}$; (ii) minimum consumption per
predator $C$ with range: 500-700 voles yr$^{-1}$ weasel$^{-1}$;
(iii) half saturation constant $D$ with range: 5-6 voles
ha$^{-1}$; (iv) predator-prey constant ratio $Q$ with range: 20-40
voles weasel$^{-1}$. The {\it generalist predation} is
characterized by: (i) the maximum rate of mortality $G$ with
range: 70 - 125 voles ha$^{-1}$ yr$^{-1}$ and (ii) half-saturation
prey density $H$ with range: 11-16 voles ha$^{-1}$.

First, we try to fit this model to the experimental data taking
into account measurements of both populations. To avoid the
problem related to the fact that continuous model is being fitted
to the experimental points measured only once a year we
interpolate experimental points for predator and prey using a
piecewise cubic Hermite interpolation with time step $h=0.001$
year. The corresponding results of the fit are shown by the dashed
black line in the Fig.~\ref{fig:lemming_data}(a) and (b). We note
that the model (\ref{eq:lemming_model}) can fit very well
experimental data. However, this agreement has a limited
statistical significance since we are fitting 30 experimental
points by a nonlinear model with 18 parameters. It turns out that
the same experimental data can be well fit in a broad range of the
values of the model parameters. A detailed study of the landscape
of the log-likelihood function is needed to choose the most
probable model. We defer this study to a future publication. In
the present study our main goal is to verify that even in the
absence of the measurements of the predator population we can
still recover both hidden dynamics of the predator and the model
parameters although with degraded accuracy.
\begin{table}
    \caption{\label{tab:Lemming_mod} Inference results for the parameters of the model
    (\ref{eq:lemming_model}) obtained by two methods: Value I were obtained assuming that both
    populations (lemming and stoat) were measured; Value II were obtained assuming that only
    lemming population were measured. Experimental points in both cases were interpolated using
    piecewise cubic Hermite interpolation with time step $h=0.001$ year.}
    \begin{center}
    \begin{tabular}{c @{\hspace{0.25in}} r @{.} l @{\hspace{0.35in}} r @{.} l}
        \hline \hline
        {\it Parameter} & \multicolumn{2}{l}{\it Value I} & \multicolumn{2}{l}{\it Value II} \\
        \hline
        $r$      & $2$&$24$     & $4$&$53$      \\
        $s$      & $0$&$76$     & $0$&$99$      \\
        $rK'/K$  & $-4$&$07$    & $-7$&$26$     \\
        $a$      & $-7$&$96$    & $-14$&$18$    \\
        $g$      & $0$&$41$     & $0$&$39$      \\
        $sQ/Q'$  & $-0$&$82$    & $-0$&$95$     \\
        $e_{11}$ & $0$&$63$     & $0$&$35$      \\
        $e_{21}$ & $0$&$32$     & $0$&$09$      \\
        $e_{12}$ & $0$&$28$     & $0$&$40$      \\
        $e_{22}$ & $-0$&$21$    & $-0$&$29$     \\
        \hline \hline
    \end{tabular}
    \end{center}
\end{table}

To this end we now infer both hidden dynamics of the stoat
population and the model parameters assuming that only population
of lemming was measured. The corresponding inference results are
shown in the Fig.~\ref{fig:lemming_data}(a) and (b) by the red
dash-dot lines. The values of the parameters inferred in both
cases are summarized in the Table~\ref{tab:Lemming_mod}.

We conclude that even in the case of incomplete corrupted by noise
measurements of the population dynamics our method allows one to
recover both hidden dynamics of invisible predator and the model
parameters.

\subsection{3D distribution of the predator trajectories}
 \label{s:trajectories}

\begin{figure}[ht!]
\begin{center}
\includegraphics[width=10cm,height=10cm]{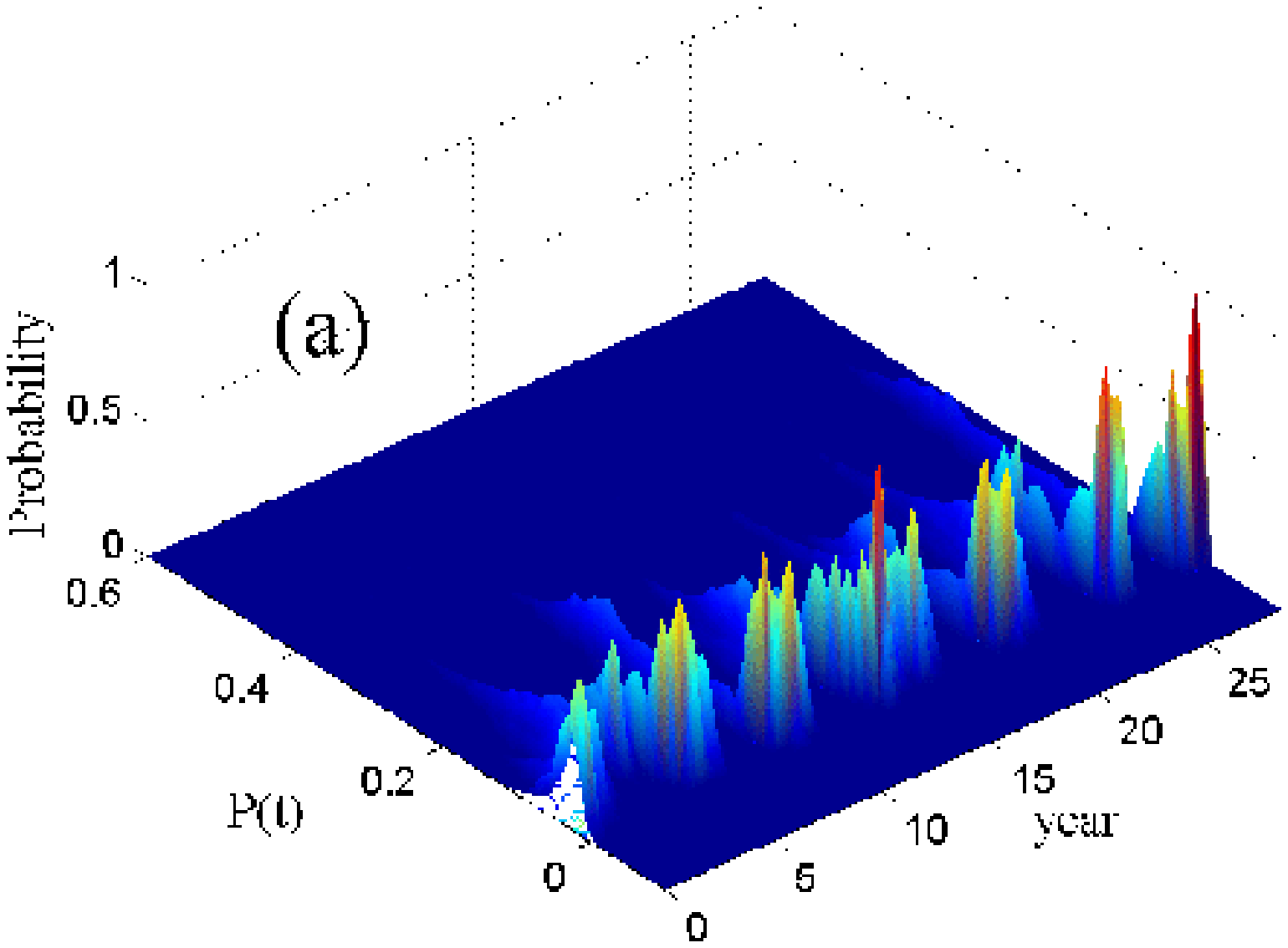}

\includegraphics[width=10cm,height=5cm]{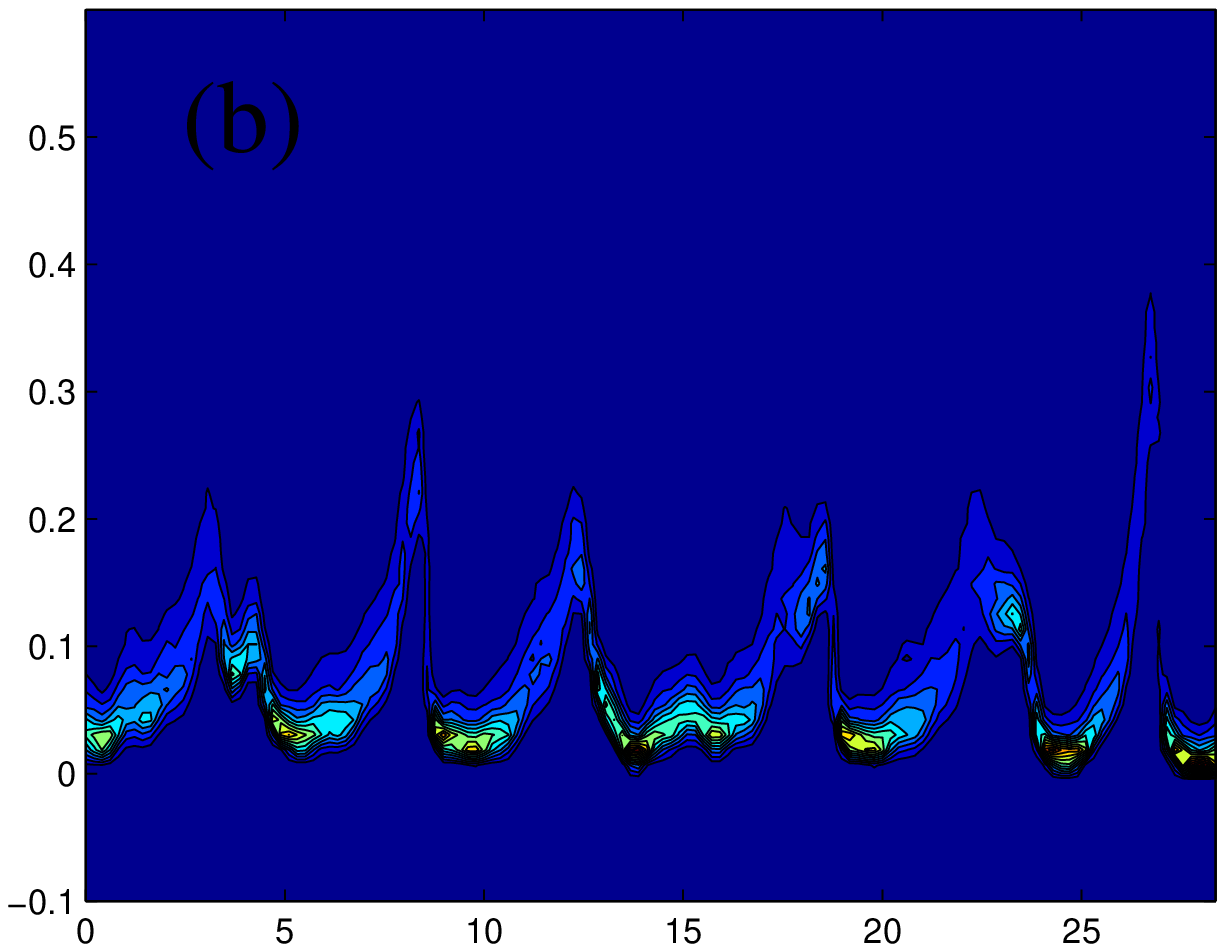}
\end{center} \caption{\label{fig:3dtrajectory} (a) probaility distribution of the
predator trajectories at local minima of LPDF; (b) the
corresponding contour plot. Local minima of LPDF with respect to
four model parameters were sampled in the region: $r=\{5.2000\pm
4, rK'/K=-5.2000\pm 4, s=1.2000\pm 1.1, sQ/Q'=-1.2000 \pm1.1\}$.
Other parameters in this test were: $C/Q'=20.9223$, $G/K'=0.2401$,
$re_1=2.2493$, $se_2=0.5027$, $(H/K')^2=0.04$, $D/K'=0.04$.}
\end{figure}

 Finally we analyze a distribution of the
 most probable predator trajectories for different model parameter values taken at multiple minima
 of LPDF discussed above (see Fig.~\ref{fig:vole_rs_dist}).
 We search for local minima of LPDF with respect to the set of coefficients ${\cal
 M}=\{r, rK'/K, s, sQ/Q'\}$ in the region $\{5.2000\pm 4,-5.2000\pm 4,
1.2000\pm 1.1,-1.2000
 \pm1.1\}$.
 At each local minima ${\cal M^{\prime}}$ we find the most probable predator trajectory $\bx_{\rm opt}(t)$ by solving a
 boundary value problem
 described above and attach the statistical weight to this trajectory
 $\propto\exp(-S(\bx_{\rm opt}(t),{\cal M^{\prime}})$. The
 resulting 3D distribution of the weighted predator
 trajectories is shown in the Fig \ref{fig:3dtrajectory}.

\section{\label{sec:MCMC}
Comparison of path-integral based inference with Markov Chain
Monte Carlo method}

To compare directly the results of the reconstruction obtained by
the path-integral method and by MC algorithm we simplify problem
and consider oscillations in a two-dimensional system of the form
\begin{eqnarray}
  \label{eq:2D_system}
\begin{array}{ll}
    &\dot x_1 = 1.5x_2 + x_1^2x_2 - 0.2x_1^3 + \sqrt{D_{11}}\xi_1(t), \\
    &\dot x_2 = - x_1 + \epsilon(1-x_1^2)x_2 +
    \sqrt{D_{22}}\xi_2(t),\\
\end{array}%
\end{eqnarray}
where $\epsilon=0.1$ and $D_{11}=D_{22}=0.04$. We assume that only
$x_1(t)$ is measured with measurement noise $\eta(t)$ of intensity
$N=0.2$ to produce an observed time series
\[y_1(t)=x_1(t)+\sqrt{N}\eta(t),\]
while the second variable is missing.
\begin{figure}[ht!]
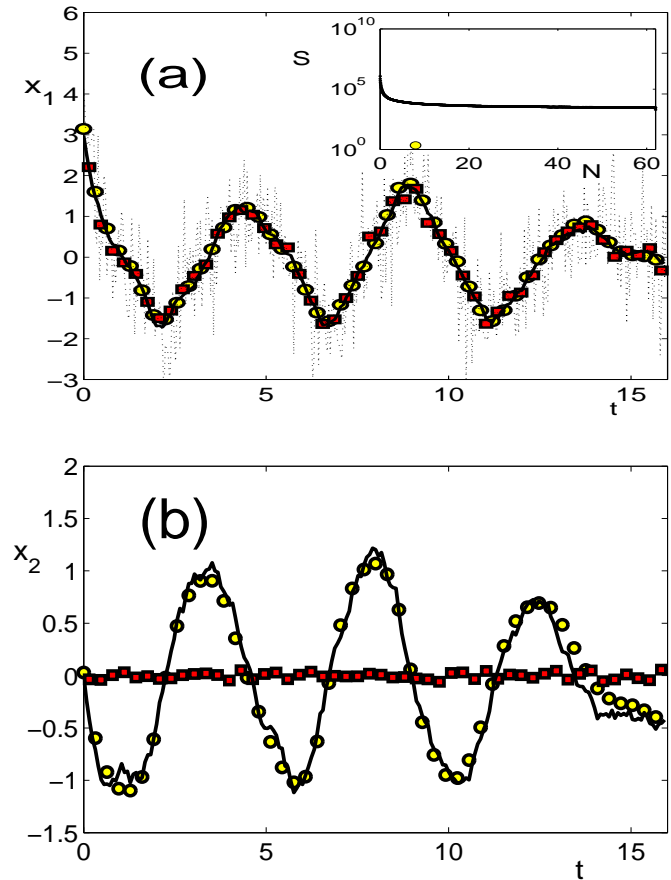

\begin{center}
\includegraphics[width=10cm,height=6cm]{bvp_mcmc_1.eps}

\includegraphics[width=10cm,height=6cm]{bvp_mcmc_2.eps}
\end{center} \caption{\label{fig:bvp_mcmc} Result of the direct comparison
of the path-integral and MCMC techniques for (a) observed variable
$x_1(t)$ and (b) hidden variable $x_2(t)$. The actual dynamical
trajectories $x_1(t)$ and $x_2(t)$ are shown by solid lines. The
measured trajectory $y_1(t)$ is shown by dashed black line in
figure (a) and is taken as initial guess for the solution
$x_1(t)$. For an unobservable trajectory $x_2(t)$ initial guess is
taken to be $y_2(t)=0$. The solution of the boundary value problem
is shown by yellow circles. The MCMC solution is shown by red
squares. The inset in the figure (a) shows the variation of the
cost function as a function of time for MCMC algorithm (black
dots) and for boundary value method (yellow circle).}
\end{figure}\noindent
We find maximum of the posterior PDF $p_{\rm ps}(\bx,\CM|\by)$ in
the space of dynamical trajectories $\{\bx(t)\}$ by applying two
methods: path-integral approach as described above and Markov
Chain Mote Carlo (MCMC) using Metropolis-Hastings algorithm within
Gibbs sampling scheme (see e.g.~\cite{Ruanaidh:96}). The results
are shown in the Fig. 4. It can be seen from the figure that the
MCMC algorithm can indeed be used to reconstruct dynamical
trajectory from the noisy measurements. However, in the case of
missing variable the MCMC fails to recover correct solution. The
reason is that the later requires large smooth variations of the
trajectory, while the MCMC algorithm is searching in the space of
discontinuous nondifferentiable trajectories and as a result
converges to multiple deep spurious minima produced by the terms
of the order $\frac{(x_{k+1}-x_{k})^2}{h^2}$ in the cost function
(\ref{eq:action}). Similar problem appears already in
deterministic case~\cite{Kurths:04}, where the multiple shooting
technique is applied to solve the problem. We note that our
approach is more general. It is valid both in stochastic and
deterministic case and avoids logistic and technical problems
related to dividing trajectory on arbitrary number of piece-wise
continuous solutions and on gluing these solution together.

\subsection{Lorenz attarctor \label{ss:Lorenz}}
\begin{figure}
 \includegraphics[width=4in,height=4.5in]{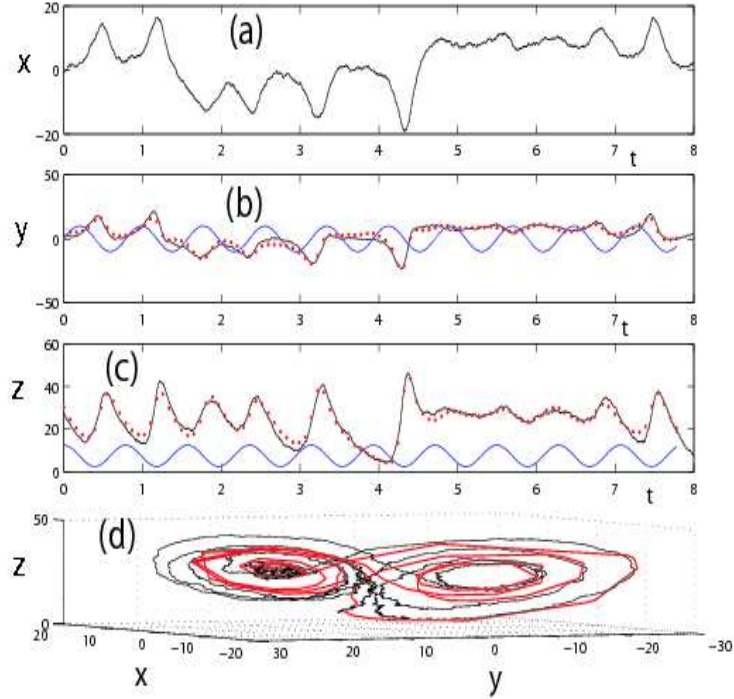}\vspace{-0.6in}
  \caption{\label{fig:Lorenz} (a) Measured variable $x(t)$ for the following system parameters:
  $D_{11} = 2.5$, $D_{22} = 3.0$, and $D_{33} = 3.5$, $r=28$, $\sigma=10$, $b=8/3$.
  (b) and (c) Actual values of the unknown dynamical variables $y(t)$ and $z(t)$ are shown by the black line.
  Inferred values are shown by the red dots. (d) Actual trajectory of the system in 3D space of the variables
  $x, y, z$ (black solid line) is compared with the inferred trajectory (red line).}
\end{figure}

WE found our method to be sufficiently robust to work in the case
of more then one hidden variable. To demonstrate this we consider
 the archetypical chaotic nonlinear system of Lorenz,
\begin{equation}
    \label{eq:Lorenz}
    \begin{array}{rcl}
    {\dot x}_{1} & = & \sigma \, (x_{2} - x_{1}) + \xi_1(t), \\
    {\dot x}_{2} & = & r \, x_{1} - x_{2} - x_{1} \, x_{3} + \xi_2(t), \\
    {\dot x}_{3} & = & x_{1} \, x_{2} - b \, x_{3} + \xi_3(t),
    \end{array}
\end{equation}
driven by  zero-mean white Gaussian noise processes $\xi_{l}(t)$
with covariance $\langle \xi_{l}(t) \, \xi_{l'}(t') \rangle = D_{l
l'} \, \delta(t - t')$.  Synthetic data (with no measurement
noise) were generated by simulating (\ref{eq:Lorenz}) using the
standard parameter set $\sigma = 10$, $r = 28$, $b = \frac{8}{3}$,
and for various levels of dynamical noise intensities. It is
assumed that the trajectory component $x(t)$ shown in
Fig.~\ref{fig:Lorenz}(a) is measured directly (no measurement
noise) while the components $y(t)$ and $z(t)$ are not observed
(hidden variables). The results of the trajectory inference  are
shown in Fig.~\ref{fig:Lorenz}(b,c,d).
\noindent
\section{Conclusions}
 \label{s:conclusions}

Is has been assumed up to now that a lack of observational data
for predator populations constituted a fundamental obstacle to the
inference of ecological parameters from  experimental
data~\cite{Hanski:97,Turchin:00,Hanski:01,Hanski:03}. A conclusion
that can be drawn from the above results that this is not
necessarily the case.  Using the methods described above it is
possible to reconstruct both invisible dynamics of predators
population and the model parameters directly from measurements of
prey populations, even those containing some measurement errors
and uncertainties.

Note, that much of the studies across many scientific disciplines
rely on the analysis of the extremal properties of the effective
action similar to (\ref{eq:action}) in various function spaces
(cf.
e.g.~\cite{Borland:92a,Borland:96,Luchinsky:97c,Smelyanskiy:97a,Luchinsky:02b,Smelyanskiy:05b}).
For example, solution of the problems of large occasional
deviations   in noisy dynamical systems is given by the minimum of
the functional (\ref{eq:action}) with no measurement term. Unlike
the dynamical inference  searching for the trajectory and model
parameters that the system has with high probability,
  the theory of large
 deviations is concerned with an optimal fluctuation,  or a   \emph{least improbable path} of
the system to reach a remote state from the attractor during the
rare event.  However the solution of both problems provides a
global minimum to the functional (\ref{eq:action}) in the space of
dynamical trajectories and is equivalent to a certain  Hamiltonian
dynamics in an extended phase space. In the theory of large
deviations the corresponding Hamiltonian is called
Wentzel-Friedlin Hamiltonian~\cite{Freidlin:84a}. The dynamical
quantities appearing within this Hamiltonian theory  have precise
physical meaning and are accessible for direct experimental
measurements~\cite{Luchinsky:97c,Dykman:00a,Luchinsky:02b}. Very
similar optimization problems also occur in the context of
stochastic optimal control of large deviations
~\cite{Freidlin:70b,Smelyanskiy:97a,Luchinsky:02b} and the related
Hamiltonian can sometimes be identified~\cite{Luchinsky:02b} with
so-called Pontryagin Hamiltonian~\cite{Hagedorn:82} playing a key
role in the theory of optimal control. We note however that the
dynamical inference Hamiltonian $H_\CY(\bx,\bp)$
(\ref{eq:the_Hamiltonian}) is of a qualitatively new type. It
depends explicitly on the time-varying measurement signal $\by(t)$
that plays a role of a 'control force' in  the Hamiltonian
dynamics. These considerations suggest that the proposed
path-integral approach to the problem of dynamical inference with
hidden variables is a general one and sets the solution of this
problem into the standard mathematical context. It is valid both
in deterministic and stochastic case and is a natural
generalization of the earlier {\it ad hoc} approach to the
dynamical inference of deterministic
systems~\cite{Domselaar:75,Kurths:04}.

We believe that methods of Hamiltonian theory will provide a new
topological insight to the solutions of complex  problems of a
dynamical inference with hidden variables. For example, in many
cases the observed data are not sufficient to discriminate with
high probability between the different values of system parameters
and/or the forms of its hidden trajectory component. This
corresponds to a certain  'degeneracy' set in the joint functional
space $(\bx(t),\CM)$ where the functional $S$  takes a constant
maximum value. In general, the degeneracy set will be determined
by the properties of the corresponding Lagrangian manifold
associated with the auxiliary Hamiltonian system
(\ref{eq:the_Hamiltonian}) and conditions $\bp(0)=\bp(T)=0$. We
also note that whenever the dynamical inference converges to a
right solution  the inferred system trajectory and parameter
values correspond to a sufficiently small momentum $|\bp(t)|$ (of
the order of noise intensities, $D,\, N$) and the minimum action
$S^{\rm opt}\sim 1$. However in certain cases the global minimum
of $S$ corresponds to a much larger momentum $|\bp(t)|\gg N,\,D$
and $S^{\rm opt}\gg 1$. Then  the fitness to the data $\CY$ is
poor for \emph{any} choice of parameters and trajectory. This
implies that model assumptions (\ref{eq:SDE}) do not capture some
important properties of the real-world system (a so called, \lq\lq
model error"). Overall, the locations  of maxima of effective
action $S$ dominating the LPDF, their relative weights, as well as
the topological structure  in the joint functional space
$(\bx(t),\CM)$ answer the statistical question of what \emph{can}
or \emph{cannot} be learned with a high likelihood about the
system at hand given the available data and basic assumptions
about the dynamical model.

Our results also reveal a remarkable property of the dynamical
inference with incomplete measurements. In the absence of the
model error the system parameters can be learned with
uncertainties $\langle (\delta M_{\alpha})^2\rangle$
 that are \emph{not} limited by the dynamical nor
measurement noise intensities. In particular, $\langle (\delta
M_{\alpha})^2\rangle \lesssim 1/T$, for large $T$ (see Appendix
for the details of the derivation). On the other hand, the
uncertainty in the inferred system trajectory $\langle (\delta
x_j(t))^2\rangle$ is bounded from below by the dynamical and
measurement noise. This effect can appear counterintuitive to a
reader, because hidden variables and model parameters are trading
against each other in the log-likelihood (\ref{eq:action}) that
could seemingly cause the parameter and trajectory errors to be
comparable with each other.
 The explanation for
the above effect is that the trajectory  points $\bx(t_m)$ at
closely spaced instances of time $t_m$ are correlated with each
other, those correlations are being extracted and accumulated
during the dynamical inference which we presented in the paper and
this leads to the shrinking of the parameter error with time below
the noise level.

The proposed method should be applicable to a broad range of
problems in science and technology ranging from extracting
parameters of molecular motors from the measurements of their
progression along microtubules~\cite{Visscher:99,Kawaguchi:01} to
the inference of a climate forcing mechanisms from reconstructed
from the measurements of carbon dioxide in ocean
sediment~\cite{Rahmstorf:02}. We also expect this method to be
particularly useful in the context of physiological measurements
where it is especially important to relate difficult-to-access
parameters to noninvasively-measured
data~\cite{Seidel:95,Seidel:98a}. The open question to be
addressed in the near future is an extension of this theory to
quantum and spatially extended systems.

\section*{APPENDIX: BAYESIAN INFERENCE OF CONTINUOUS NOISE-DRIVEN
DYNAMICAL SYSTEMS  FROM INCOMPLETE MEASUREMENTS }

Within the Bayesian framework the problem of dynamical inference
is to determine the conditional  probability density functional
(PDF) defined over the set of the unknown quantities $(\bf
x(t),\CM)$, subject to observations $\by(t)$. The later, so-called
{\it posterior} PDF, $p_{\rm{ps}}[\bx(t);\CM|\,\by(t)]$ is found
using Bayes' theorem
\begin{equation}
    \label{eq:Bayes}
    p_{\rm{ps}}[\bx(t);\CM|\,\CY] \propto \,p_{\rm ob}[\by(t)\,|\,\bx(t),\CM] \,
    p_{\rm{pr}}[\bx(t),\CM].
\end{equation}
Here the missing proportionality coefficient is simply a
normalization factor. $p_{\rm{pr}}[\bx(t),\CM]$ is a so-called
{\it prior} PDF  that provides the joint statistical information
about  $\bx(t)$ and $\CM$
 before the measurements $\by(t)$ were made. The prior
PDF can be written in the form: $p_{\rm{pr}}[\bx(t),\CM]={\cal
P}[\bx(t)|\,\CM]\,p_0(\CM)$. Here $p_0(\CM)$ is some prior
distribution of model parameters and ${\cal P}[\bx(t)|\,\CM]$ is
the PDF of finding a  realization of the dynamical trajectory
$\bx(t)$ for a given set of the system parameters $\CM$
~\cite{Ludwig:75,Graham:77a,Dykman:90}. This functional directly
depends on the form of the stochastic dynamical model
(\ref{eq:SDE}) and its parametrization. For example, in the case
of the additive white noise considered in (\ref{eq:SDE}) this
functional has the form ~\cite{Ludwig:75,Graham:77a}
\begin{eqnarray}
{\cal P}[\bx(t)|\,\CM]&\propto&
\left(\left(\frac{2\pi}{h}\right)^L {\rm det}\hat
\bD\right)^{-\CK/2}\,\label{paths}\\
&&\times\,\exp\left[-\frac{1}{2}\int_{0}^{T}dt\left( \bnabla\cdot
\bK +(\dot\bx-\bK^T)\hat
 {\bf D}^{-1}(\dot\bx-\bK)\right)\right],\label{paths}
 \end{eqnarray}
 \noindent
where $\bx\equiv \bx(t)$, $\bK\equiv \bK(\bx(t),\bC)$ and a
coefficient of proportionality is a normalization factor.

In Eq.(\ref{eq:Bayes})  $p_{\rm ob}[{\cal Y}|\bx(t),\CM]$ is a
conditional PDF to observe the measurement signal $\by(t)$ for a
specific realization of a system trajectory $\bx(t)$ and model
parameters $\CM$. For the continuous-time measurement model
considered in (\ref{eq:SDE}) this PDF takes the form
\begin{eqnarray}
p_{\rm ob}[{\cal
Y}\,|\bx(t),\CM]&=&\left(\left(\frac{2\pi}{h}\right)^M {\rm
det}\hat
\bN\right)^{-\CK/2}\,\label{Pobs}\\
&&\times\,\exp\left[-\frac{1}{2}\int_{0}^{T}dt\, \left(\by(t)-\hat
\bB \,\bx(t) \right)^T{\hat
 \bN}^{-1}\left(\by(t)-\hat \bB\,\bx(t) \right)\right],\nonumber\end{eqnarray}
\noindent and describes the zero-mean Gaussian statistics of the
measurement error $\bbeta(t)=\by(t)-\hat \bB\,\bx(t)$. Returning
back to the original discreet-time measurements
$\CY=\{\by(t_m),\,t_m=m\,h,\, m=1:\CK\}$ one gets
$\langle\beta_{k}(t_m)\beta_{k'}(t_m')\rangle=N_{kk'}/h\,\delta_{mm'}$.

Prior PDF $p_0[\CM]$  usually represents  a posterior PDF obtained
as a result of the  dynamical inference based on the
\emph{previous}  sets of data and on the expert knowledge about
possible domains for  the system parameters. Often  the inference
is entirely based on the present
 set of data and the prior PDF is assumed to be completely
 uniform.
In this case  the posterior PDF $p_{\rm{ps}}[\bx(t);\CM|\,\CY]$ in
 (\ref{eq:Bayes}) is usually referred to as a  likelihood PDF.
 We denote the later as $p_{\CY}[\bx(t);\CM]$ and obtain:
 \begin{equation} p_{\CY}[\,\bx(t);\CM]\propto p_{\rm ob}[{\cal
Y}\,|\bx(t),\CM]\,{\cal P}[\bx(t)|\,\CM].\label{pY}
 \end{equation}
\noindent Using Eqs.~(\ref{paths}) and (\ref{Pobs}) in (\ref{pY})
one can rewrite the likelihood PDF in the form
\begin{equation}
p_{\CY}[\bx(t);\CM]=A_\CY\, \exp\left(-
S_\CY[\bx(t);\CM]\right).\label{like}
\end{equation}
\noindent where the negative log-likelihood function $S_\CY$ is
given in (\ref{eq:action}) and $A_\CY$ is a normalization factor
that does not depend on $\bx(t)$ nor $\CM$.

\paragraph{Calculation of the expectation values using  the
maximum-likelihood estimation.} In the asymptotic limit of a
sufficiently long and dense data record $\CY$
 and low noise intensities the PDF
$p_{\CY}[\,\bx(t);\CM]$ is, generally, a very steep function of
its arguments and the derivatives of $S_\CY$ with respect to
$\bx(t)$ and $\CM$ are much greater then 1 (assuming all
quantities are dimensionless). In this case the expectation values
of the system trajectory $\langle \bx(t)\rangle$ and the model
parameters $\{\langle \CM_\alpha\rangle\}$ for a given measurement
record $\CY$ can be obtained by computing a maximum of the
likelihood PDF in the joint space $(\bf x(t); \CM)$. The
conditions for the  maximum  have the form of the  variational
equations
\begin{equation}
\frac{\delta S_\CY}{\delta \bx(t)}=0,\label{cond-x}\\
\end{equation}
\noindent that have to be solved simultaneously with the system of
the algebraic equations
\begin{equation}
\frac{\partial S_\CY}{\partial \bc}=0,\label{cond-c}
\end{equation}
\noindent\vspace{-0.2in} \begin{equation} \frac{\partial
S_\CY}{\partial \hat \bD}=0,\qquad \hat \bD=\hat
\bD^T,\label{cond-D}\end{equation}\noindent \vspace{-0.2in}
\begin{equation}
\frac{\partial S_\CY}{\partial \hat \bN}=0,\qquad
\bN=\bN^T,\label{cond-N}\end{equation}\noindent\vspace{-0.2in}
\begin{equation}\frac{\partial S_\CY} {\partial \hat
\bB}=0.\label{cond-B}\end{equation} \noindent Eqs.(\ref{cond-x})
for the minimum of the action $S_\CY$ with respect to the
trajectory components $x_i(t)$  correspond to the Hamiltonian
equations (\ref{eq:the_Hamiltonian}),(\ref{eq:HE}) with the
appropriate
 boundary conditions described in the main
text.

 Inference of the model parameters was considered in
\cite{Smelyanskiy:05b} under the simplifying assumptions that the
measurement noise is zero, there are no hidden variables and the
force field is linear in the parameters $\{c_\alpha\}$ (but
generally, nonlinear in $\bx$). Below we provide the
generalization of the results of the Ref.~\cite{Smelyanskiy:05b}
that allows us to infer the unknown parameters of the measurement
model and does not relay on  the linearity of $\bK$ in $\bc$.

The  Eq.~(\ref{cond-c}) gives the conditions of the minimum of
$S_\CY$ with respect to the parameters $\{c_\alpha\}$ of the force
field $\bK(\bx(t),\bc)$. Using the Eq.~(\ref{eq:action}) we obtain
these conditions in the following form:
\begin{equation}
\int_{0}^{T}dt\,\frac{\partial \bK}{\partial
c_\alpha}\,\hat\bD^{-1}\,[\dot
\bx(t)-\bK(\bx(t),\bc)]=\frac{1}{2}\int_{0}^{T}dt\frac{\partial
}{\partial c_j}\,\bnabla\cdot\bK(\bx(t),\bc).\label{c}
\end{equation}
\noindent Solving the Eqs.~(\ref{cond-D}) and (\ref{cond-N}) with
respect to $\hat\bD$ and $\hat \bN$, respectively, we obtain
\begin{eqnarray}
D_{ij}&=&\frac{1}{\CK}\int_{0}^{T}dt\,[\dot
x_{i}-K_{i}(\bx(t),\bc)]\,[\dot x_{j}-K_{j}(\bx(t),\bc)],\label{D}\\
N_{kl}&=&\frac{1}{\CK}\int_{0}^{T}dt\,[\dot
y_{k}(t)-\sum_{i=1}^{L}B_{ki}x_{i}(t)]\,[y_{l}(t)-\sum_{j=1}^{L}B_{lj}x_{j}(t)]\label{N}.
\end{eqnarray}
\noindent Finally, the Eq.~(\ref{cond-B}) can be re-written in the
explicit form of the system of linear equations for the matrix
elements of $\hat \bB$
\begin{equation}
\sum_{k,i}\Lambda_{ki}^{k'i'} \,B_{k'i'}=W_{ki},\label{B}
\end{equation}
\noindent where
 \begin{eqnarray}
&&\Lambda_{ki}^{k'i'}=(\hat
\bN^{-1})_{kk'}\,\int_{0}^{T}dt\, x_{i'}(t)\,x_{i}(t),\label{Lambda}\\
&&W_{ki}=\sum_{l}(\hat
\bN^{-1})_{kl}\int_{0}^{T}dt\,y_{l}(t)\,x_{i}(t).\label{W}
\end{eqnarray}
\noindent One solves simultaneously Eqs.~(\ref{c})-(\ref{W}) and
the Hamiltonian equations (\ref{eq:the_Hamiltonian}),(\ref{eq:HE})
and selects the solution with the minimum value of $S_\CY$.

\paragraph{Calculation of the variances.}
 We now consider how the variances of the  model parameters
around the maximum of the LPDF depend on the noise intensity and
length of the observation record. We  focus on demonstrating of
the main effect mentioned in Conclusion and for brevity we assume
that there are two dynamical variables, one of them, $x_1$, is
hidden,  and the other, $x_2$, is observed with zero measurement
error, $x_2(t)=y_2(t)$. We assume that the correlation matrix
$\hat \bD$ of the dynamical noise is diagonal with the small
nonzero matrix elements $D_{j}\equiv D_{jj}\ll 1$. We  also assume
that there is only one unknown model parameter $c$ and it enters
the expression for the vectorial force field $\bK=\bK(\bx,c)$.

The action functional in the reduced space
$S_{\CY}[x_1(t),y_2(t);c]\equiv s[x_1(t);c]$ has the form
\begin{eqnarray}
s[x_1(t);c]&=&\frac{1}{D_{1}}\int_{0}^{T}dt\,
\left[\frac{1}{2}(\dot
x_1(t)- \bar K_{1}(x_1(t),y_2(t);c))^2-V(x_1(t),t,c)\right]\\
V(x_1,t,c)&\equiv&-\frac{D_{1}}{2 D_{2}}(\dot
y_{2}(t)-K_{2}(x_1,y_2(t);c))-\frac{D_1}{2}\,\sum_{j=1}^{2}\frac{\partial
K_{j}(x_1,y_2(t);c)}{\partial x_j}.\label{Sbar}
\end{eqnarray} \noindent
At certain point $(x_{1}^{\rm opt}(t), c^{\rm opt})$ where the
action $ s[x_1(t),c]$ reaches its minimum  the conditions $\delta
s/\delta x_1(t)=0$ and $\partial s/\partial c=0$ are satisfied.
Consider now the trajectory $x_1=x(t,c)$ corresponding to the
\emph{partial} minimum of the action with respect to $x_1(t)$ with
the value of the model parameter fixed: $\min_{x_1(t)}
s[x_1(t),c]=s[x(t,c),c]$. The
 Hamiltonian equations (\ref{eq:HE}) for $x(t,c)$ have
the following form:
\begin{eqnarray}
\dot x_1&=&p_1+K_1(x_1(t),y_2(t);c),\\ \dot p_1&=&-p_1
\frac{\partial K_1(x_1(t),y_2(t); c)}{\partial x_1}-\frac{\partial
V(x_1(t),t; c)}{\partial x_1}, \quad
p_1(0)=p_1(T)=0.\label{Hambar}
\end{eqnarray}
\noindent Of central interest for us here is the coefficient of
expansion of the action $s[x(t,c);c]$ in
 $c-c^{\rm opt}$
 \begin{equation} s[x(t,c),c]\approx
\frac{a}{2 }(c-c^{\rm opt})^2,\quad \langle \delta c^2\rangle
=a^{-1},\label{a}
\end{equation}
\noindent  that equals to the inverse variance of the model
parameter $c$. To calculate this coefficient we expend the
trajectory $x(t,c)\approx  (c-c^{\rm opt})\,\xi(t)+{\cal
O}\left((c-c^{\rm opt})^2\right)$.  Then, using
(\ref{Sbar}),(\ref{Hambar}) we obtain in the leading order in
$D_{11},D_{22} \ll 1$
\begin{equation}
a\equiv
a(T)=\int_{0}^{T}\left[\frac{1}{2D_1}\left(\dot\xi(t)-\frac{\partial
\bar K_{1}}{\partial c}-\xi(t)\frac{\partial \bar K_{1}}{\partial
x_1}\right)^2+\frac{1}{2D_2}\left(\frac{\partial \bar
K_{2}}{\partial c}+\xi(t)\frac{\partial \bar K_{2}}{\partial
x_1}\right)^2\right].\label{aa}
\end{equation}
\noindent The function $\xi(t)$ can be obtained from solution of
the following system of  equations obtained by linearization of
 equations (\ref{Hambar}) around the Hamiltonian
trajectory $(x^{\rm opt}(t),p^{\rm opt}(t))$ in the extended space
($x,p)$ corresponding to the full minimum of the action
$s[x(t;c),c]$:
\begin{eqnarray}
\dot\eta(t)&=&-\eta(t)\frac{\partial K_1^{\rm opt}}{\partial
x_1}-\xi(t)\,p^{\rm opt}(t)\frac{\partial^2 K_1^{\rm
opt}}{\partial x_{1}^{2}}-p^{\rm opt}(t)\frac{\partial^2 K_1^{\rm
opt}}{\partial x_{1}\partial
c}\nonumber\\
&&-\xi(t)\frac{D_1}{D_2}\,\left(\frac{\partial K_1^{\rm
opt}}{\partial x_{1}}\right)^2-\frac{D_1}{D_2}\,\frac{\partial
K_1^{\rm opt}}{\partial x_{1}}\,\frac{\partial K_1^{\rm
opt}}{\partial c},\label{deta}\\
\dot \xi(t)&=&\eta(t)+\xi(t)\,\frac{\partial K_1^{\rm
opt}}{\partial x_1}+\frac{\partial K_1^{\rm opt}}{\partial
c},\label{dxi}\\
\eta(0)&=&\eta(T)=0,\label{boundary}\\
\xi(t)&\equiv&\frac{\partial x(t;c^{\rm opt})}{\partial c},\quad
\eta(t)\equiv\frac{\partial p(t;c^{\rm opt})}{\partial c},\quad
K_{1}^{\rm opt}\equiv K_1(x_{1}^{\rm opt},y_2(t),c^{\rm
opt}).\nonumber
\end{eqnarray}
\noindent (all the partial derivatives of $K_1$ above are
evaluated at the arguments $x_1(t)=x^{\rm opt}(t)$ and $c=c^{\rm
opt}$).

We note that the integrand in the expression for $a(T)$ represents
a sum of squares and therefore $a(T)$ is a growing function of
$T$, implying that the variance $\langle \delta c^2\rangle$
shrinks down with $T$. Assume now that the measurement of the
trajectory component  $y_{2}(t)$ varies periodically for large $t$
(system approaches a periodic attractor). This variation will play
a role of a periodic forcing in Eqs.~(\ref{deta})-(\ref{boundary})
and the long-time solutions of those equations, $\xi(t), \eta(t)$
will also have a periodic component. That means that $a(T)$ in
(\ref{aa}) is growing at least linearly with $T$, an assertion
made in the Conclusion. Dynamical inference with hidden variables
in systems with chaotic attractors will be considered elsewhere.

\end{document}